# Theory of the ultra-intense short-pulse laser interaction with under-dense plasma

J. Yazdanpanah[1*], E. Yazdani[2], A. Chakhmachi[1] and E. Khalilzadeh[1, 3]

[1]*The Plasma Physics and Fusion Research School, Tehran, Iran*

[2] *Laser and optics research school, P.O. Box 11365-3486, Tehran, Iran*

[3]*Department of Physics, Kharazmi University, 49 Mofateh Ave, Tehran, Iran*

**Abstract**

A comprehensive theory is proposed to describe the propagation and absorption of ultra-intense, short laser pulse through the under-dense plasma. The kinetic aspects of plasma are fully incorporated using extensive particle-in-cell (PIC) simulations. It is turned out that the plasma behavior is characterized by both its density and the ratio of the pulse length to the plasma wavelength. According to exact analyses and direct simulation evidences, at ultra-low densities the laser pulse is adiabatically depleted (absorbed) by the wake excitation. And the depletion is accompanied by the overall radiation red-shift. At these densities, for pulse lengths larger than the plasma wavelength the Raman type scatterings also occur without causing instability. When the plasma density grows toward the critical density, a completely new regime appears with the main character of highly unsteady light propagation. Here, based on analyses and simulations, the radiation pressure induced wave breaking (RPIWB) heavily destroys the oscillatory structure of the electron wave behind the first plasma period. On the other hand, the electron density profile induced by the ponderomotive force ahead of the pulse begins to steepen and eventually acts as a flying mirror. As a result, the pulse becomes bake scattered and its penetration becomes limited. The radiation pressure of reflecting pulse sustains a longitudinal electric field which accelerates electrons produced via RPIWB causing the plasma to undergo volumetric heating. Mostly important, based on a newly proposed model it is shown that the interaction becomes saturated at a definite time and that the overall absorption and plasma heating are directly related to this time. Also it is found that the saturation time decreases by a factor of $n_{e0}/\gamma_{\perp 0} n_c$ ($n_{e0}$, $n_c$ and $\gamma_{\perp 0}$ are initial electron density, critical density and initial laser gamma factor). Other details, mainly related to different active electron acceleration mechanisms, have minor effects in absorption which are also discussed.

______________________

[*]jamyazdan@gmail.com

## I. INTRODUCTION

The current ability of production of ultra-short, ultra-intense laser pulses [1-2] have stimulated extensive researches in the field of laser-matter interactions. Attempts are mainly conducted toward the construction of compact particle accelerators [3-18] and bright X-ray sources [19-20], and realization of conditions for the Fast Ignition (FI) scheme of Inertial Confinement Fusion (ICF) [21-23]. The propagation of intense laser pulse through under-dense plasma is an important common part of all these scenarios. This subject has recently received a vast number of attentions [14-18, 24-37]. The sub-critical densities are either produced due to the unavoidable prepulse effects in the laser-solid interaction [13] or initially exist in the form of gas jets [28] or the state-of-the-art foam targets [25]. From the application point of view, it is mostly desired to know how the intense laser light is absorbed and scattered during its propagation through the under-dense plasma. In the laser-solid experiments, it is intended to know what amount of laser light is absorbed in the preplasma profile induced by the prepulse in the front of target before it arrives at the critical surface where a new interaction mechanism is set. In the FI scheme of ICF the crucial question is the amount of laser penetration toward the fuel core. This issue is closely related to the light absorption and scattering in the sub-critical fuel corona. In the recent experiments utilizing foam target for generation of intense particle currents, the absorption efficiency is the most important quantity. This is while; despite to these extensive quests, the theory of light propagation through the under-dense plasma has not yet been adequately developed to answer these questions. Only few theoretical works exist which are directly related to the subject [38, 39]. And to these we must add few recent experimental studies that have used foam targets to characterize the laser propagation through the transparent plasma [16, 17].

The traditional weakly nonlinear theories of light evolutions in plasma are based on the Raman and Brillouin scatterings [40-43]. Though elucidating, they are not directly applicable to relativistic intensities. This problem is mainly due to the excitation of strong space charge mode, the so called wake-field [3-6], by the intense laser pulse. This behavior is contrary to the Raman and Brillouin scatterings which suppose the space charge to grow from the noise level and to remain always small. The strong dispersion modification of plasma by the wake-field and its strength itself interplay to cause the plasma to reveal notable kinetic behaviors which eventually change the nature of light scattering into a completely new type.

The purpose of this paper is to give a comprehensive description of the laser light evolution in the under-dense plasma. This study is based on analyses performed in one spatial and three velocity dimensions (1D3V) assisted by the extensive fully kinetic particle-in-cell (PIC) simulations. The key issues of this theory are the light scattering and absorption. Trivially, the absorption is related to the global plasma heating. The absorption of ultra-short intense laser pulse in the ultra-low density plasma is investigated in Refs. [38, 39]. Note that for present laser parameters, ultra-low density $n_{e0}/n_c < 0.01$ corresponds to $L_p \leq \lambda_p^{NL}$. Here $n_c = \varepsilon_0 m_e \omega_0^2 / e^2$ is the critical density and $\lambda_p^{NL} = Fc/\omega_p$ is the nonlinear plasma wavelength. $\varepsilon_0$, $m_e$, $e$, $c$, $\omega_0$ and $\omega_p = (n_{e0} e^2 / \varepsilon_0 m_e)$ are vacuum permittivity, electron mass, electron charge, light speed, laser carrier frequency and plasma frequency respectively. $F$ is a relativistic factor depends on the dimensionless laser amplitude $a_0$, i.e. $F = 2\pi$ for $a_0 \ll 1$ and $F = 4\sqrt{1+a_0^2}$ for $a_0 \gg 1$. Ref. [38] indicates that the pulse depletion in this regime is adiabatic with being characterized by the radiation red-shift. And the pulse edge evolutions is investigated in Ref. [39]. In this study, we firstly give a more complete description of the adiabatic pulse evolutions in the ultra-low density regime along with direct verifying evidences from PIC simulations. We also show that for pulse lengths larger than the plasma wavelength the Raman type scatterings (light scattering off electron plasma wave) occur but without instability. In the other words, this scattering does not play remarkable role in pulse depletion. The limitations of the adiabatic description are line out, resulting in the identification of a new interaction regime at high densities with different characteristic behaviors. Thus, in order to complete the under-dense plasma description, we extend our investigation to plasma densities up to near critical values. It is turned out that the Radiation Pressure Induced Wave Breaking (RPIWB) heavily destroys the oscillatory structure of the electron wave beyond the first plasma period. On the other hand, the electron density profile induced by the ponderomotive force ahead of the pulse begins to steepen and eventually acts as a flying mirror. As a result, the pulse propagation becomes highly unsteady and its penetration becomes limited. The plasma undergoes volumetric heating due to phase mixing by RPIWB. The wave-break electrons are accelerated further in the longitudinal electric field sustained by the radiation pressure of the reflecting pulse. Due to the radiation reflection, the interaction becomes saturated at a definite time. This time is shown to be the most important factor in the overall absorption and plasma heating with being directly related to these quantities. It decreases by a factor of $n_{e0}/\gamma_{\perp 0}n_c$. Other details, mainly related to different active electron acceleration mechanisms, have minor effects in absorption which are also discussed. Especially, a new mechanism for direct laser acceleration is pointed out. The applicability of results in multi-dimension is commented at the end of paper.

The organization of this paper is as follow; PIC simulation parameters are summarized in section II. In section III, the basic governing equations are indicated. In section IV, the ultra-low density regime is discussed. Here, firstly the adiabatic pulse depletion for $L_P \leq \lambda_p^{NL}$ via exact analysis and direct evidences from PIC simulations is described in subsection (A). Moreover, the limitations of this theory which motivate us to investigate a different behavior in high density plasmas are deliberated. Then, the long pulse behavior $L_P > \lambda_p^{NL}$ is discussed in subsection (B). In section V. the high density regime is described using model analyses and extensive PIC simulations. This section involves three subsections. In section VI, the conclusion of the present study and the discussion of applicability of results in multi-dimensions are provided.

## II. PARAMETERS OF PIC SIMULATIONS

The simulation parameters are summarized here in order to avoid their repetition through the paper. Using PIC code which was developed by J. Yazdanpanah [44, 45], we have carried out extensive 1D3V PIC simulations over wide range of system parameters; including plasma densities from $0.01 n_c$ to $n_c$, laser intensities between $10^{18}$ and $10^{20}$ Wcm$^{-2}$ ($2 \leq a_0 \leq 10$) and laser pulse durations and pulse shapes. For most of the simulations, the laser pulse duration has been set to $\tau_L = 100 fs$, as the pulse envelop rises (with sinus functionality) within $30 fs$ to its maximum, remains constant over $40 fs$ and falling symmetrically within $30 fs$. The laser wavelength has invariantly been considered $\lambda = 1 \mu m$ in the all cases. Other simulations with different pulse duration ($\tau_L = 200 fs$, $\tau_L = 140 fs$ and $\tau_L = 80 fs$) and different pulse shapes (having different rise and fall times) have been performed. An important new aspect of our simulations is that in many cases, simulation time has been chosen in such a way that achievement of the absorption saturation be insured. This depends on plasma density and laser intensity. For all run-instances, we have used hydrogen plasma with its initial plasma profile has been invariantly step like with initial electron and ion temperatures respectively $k_B T_e / m_e c^2 = 10^{-4} (\sim 50 \text{eV})$ and $k_B T_i = 0$ ($k_B$ is Boltzmann constant). The plasma length has been chosen such that fully cover the interaction time. High spatial resolution of 200 cells per laser wavelength with at least 64 particles per cell is used in the simulations. The spatial resolution guaranties the plasma against the un-physical heating produced by the finite-grid instability, i.e. the Debye length $\lambda_D$ is appropriately resolved, $DX / \lambda_D \approx 0.3$ [44, 46]. Reflecting and open boundary conditions have been applied for particles and fields respectively.

### III.  THE FUNDAMENTAL EQUATIONS

General insights into the nature of mutual light-plasma interaction can be gained using general conservation laws which are currently well-known in the context of relativistic fluid mechanics and electrodynamics. In relativistic fluid mechanics, the fluid energy and momentum densities are components of a second-rank tensor named the energy-momentum tensor $T_q^{\mu\nu} = h_q u_q^\mu u_q^\nu / c^2 + \Theta_q^{\mu\nu}$. Here $h_q$ is the fluid internal energy-density and $u_q^\mu$ is the velocity four-vector defined as $u_q^\mu \equiv (\gamma_q c, \gamma_q \mathbf{v}_q)$ in terms of the fluid velocity $\mathbf{v}_q$ and relativistic gamma factor $\gamma_q = (1 - \mathbf{v}_q \cdot \mathbf{v}_q / c^2)^{-1/2}$. $\Theta_q^{\mu\nu}$ is a second-rank tensor which includes nonzero temperature effects [47-50]. The subscript $q$ indicates that quantities correspond to the component $q^{\text{th}}$ (electron or ion) of plasma, note that its role is different from indices which are used to describe different components of vectors and tensors. At zero temperature, we have $h_q = n_q m_q c^2 / \gamma_q$ and $\Theta_q^{\mu\nu} = \mathbf{0}$. Using $T_q^{\mu\nu}$, the fluid energy and momentum equations are compacted into the following single covariant equation,

$$\partial_\mu T_q^{\mu\nu} = F^{\nu\mu} J_{q\mu} \tag{1}$$

where $\partial^\mu = (c^{-1}\partial/\partial t, -\nabla)$ is the covariant four-derivative, $F^{\mu\nu} \equiv \partial^\mu A^\nu - \partial^\nu A^\mu$ is the electromagnetic tensor with $A^\mu = (\phi/c, \mathbf{A})$ being the potential four-vector and $J_q^\mu = Q_q n_q u_q^\mu$ is the plasma four-current with $Q_q$ and $n_q$ being elementary charge and special density of component $q$, respectively. If no special assumptions has been made on $\Theta_q^{\mu\nu}$, the $q^{th}$ plasma component at arbitrary conditions is exactly described by Eq. (1). The electromagnetic tensor $F^{\nu\mu}$ appeared in the right hand side of Eq. (1) is obtained via the solution of inhomogeneous Maxwell equations given in the covariant form by,

$$\partial_\mu F^{\mu\nu} = \mu_0 J^\nu \tag{2}$$

Here $J^\nu$ is the total four-current, i.e. $J^\nu = \sum_q J_q^\nu$. The governing equation of the light energy-momentum can be obtained from (2) [51] and takes the same form as equation (1). The so called electromagnetic energy-stress tensor $K^{\mu\nu}$ (analogous of $T^{\mu\nu}$) is defined as,

$$K^{\mu\nu} = \begin{pmatrix} u_{em} & c\mathbf{g} \\ c\mathbf{g} & -\mathbf{T}^M \end{pmatrix}, \tag{3}$$

where $u_{em} = \varepsilon_0(E^2 + c^2 B^2)/2$ and $\mathbf{g} = \varepsilon_0 \mathbf{E} \times \mathbf{B}$ are the electromagnetic energy and momentum densities, respectively. $\mathbf{E}$ and $\mathbf{B}$ are electric and magnetic fields respectively. $\mathbf{T}^M$ is the Maxwell stress tensor, with its components being given by $\mathbf{T}_{ij}^M = \varepsilon_0[E_i E_j + c^2 B_i B_j] - u_{em}\delta_{ij}$ ($\delta$ is the Kronecker delta). $K^{\mu\nu}$ is governed by [62],

$$\partial_\mu K^{\mu\nu} = -F^{\nu\mu} J_\mu. \tag{4}$$

Substituting equation (1) in the right hand of (4), we obtain the general form of local energy-momentum conservation being customarily written in terms of $q^{th}$ component,

$$\partial_\mu(K^{\mu\nu} + T_q^{\mu\nu}) = -F^{\nu\mu}\sum_{q'\neq q} J_{q'\mu}. \tag{5}$$

The particle number for component $q$ remains constant giving the continuity equation,

$$\partial_\mu J_q^\mu = 0. \tag{6}$$

Here we restrict ourselves to zero temperature for each plasma component and one spatial dimension. Through the substitution of $\nu=0$ and $\nu=1,2,3$, in Eq. (5) we respectively obtain the energy equation and different components of momentum equation for $q$ component. Considering hydrogen plasma, we obtain for energy and $x$ momentum of electrons,

$$\frac{\partial u_{EM}}{\partial t} + m_e c^2 \frac{\partial \gamma_e n_e}{\partial t} + c^2 \frac{\partial g_x}{\partial x} + m_e c^2 \frac{\partial \gamma_e n_e v_{ex}}{\partial x} = -en_i \mathbf{v}_i \cdot \mathbf{E}, \tag{7a}$$

$$\frac{\partial g_x}{\partial t} + m_e \frac{\partial \gamma_e n_e v_{ex}}{\partial t} + \varepsilon_0 \frac{\partial}{\partial x}[\frac{1}{2}E_y^2 + \frac{1}{2}c^2 B_z^2 - \frac{1}{2}E_x^2] + m_e \frac{\partial \gamma_e n_e v_{ex}^2}{\partial x} = -en_i[E_x + v_{iy} B_z] \tag{7b}$$

where subscripts $e$ and $i$ are used for electrons and ions respectively. These equations remain covariant among different reference frames. We can neglect the contribution of $v_{iy} B_z$ compared to $E_x$ in the right hand side of (7b). This is because due to their heavy mass ions remain approximately immobile in the direction of laser polarization when intensities considered here are applied. The continuity equation (6) can also be rewritten in the form,

$$\frac{\partial n_e}{\partial t} + \frac{\partial n_e v_{ex}}{\partial x} = 0. \tag{8}$$

## IV. THE ULTRA-LOW DENSITY REGIME

### A. ULTRA SHORT PULSE LENGTHS: THE ADIABATIC PULSE DEPLETION AND ITS LIMITATIONS

Let consider the nonlinear pulse depletion due to the wake excitation in the wakefield regime. Here, the pulse length is taken to be ultra-short, i.e. $L_P \leq \lambda_p^{NL}$ and plasma is taken to be very rarefied (typically having densities $n_{e0}/n_c < 0.01$) having small optical response. This subject has been loosely considered in some previous studies [38, 39]. Here a more complete description is given along with direct evidences from PIC simulations for the laser pulse evolutions. In addition, our formalism reveals the limitations of the adiabatic interpretation and helps the understanding of essential differences between the ultra-low and near critical density behaviors. To this end, let apply equations (7a), (7b) and (8) in the Pulse Co-Moving (PCM) frame which moves with velocity equal to the pulse group velocity ($v_g$) with respect to the laboratory. And assume the so-called *Quasi-Static-Approximation* (QSA) [52-55]. Based on this approximation, plasma quantities are time-independent in the PCM frame, i.e. we can set respectively in equations (7a), (7b), and (8), $\partial \gamma_e n_e / \partial t|_{PCM} = 0$, $\partial \gamma_e n_e v_{ex} / \partial t|_{PCM} = 0$ and $\partial n_e / \partial t|_{PCM} = 0$. Hereafter we use the presentation $X|_{PCM}$ to indicate that $X$ is measured in the PCM frame. For laboratory quantities no special indication will be made. From these arguments we can immediately write equation (8) as,

$$n_e v_{ex}|_{PCM} = -v_g \gamma_g n_{e0}, \quad (9)$$

with $\gamma_g = (1 - v_g^2 / c^2)^{-1/2}$ being the relativistic gamma factor attributed to $v_g$. In the right hand of Eq. (9) we have used the fact that undisturbed plasma flows with velocity $-v_g$ in the PCM frame. We have also used Lorentz transformations for density of undisturbed region to write $n_{e0}|_{PCM} = \gamma_g n_{e0}$. In the same way, under the QSA approximation equations (7a, 7b) take the form,

$$\{\frac{\partial u_{EM}}{\partial t} + \frac{\partial}{\partial x}[c^2 g_x - n_{e0} m_e c^2 \gamma_g v_g \gamma_e + n_{e0} e \gamma_g v_g \phi] = 0\}\bigg|_{PCM}, \quad (10a)$$

$$\{\frac{\partial g_x}{\partial t} + \frac{\partial}{\partial x}[\frac{\varepsilon_0}{2} E_y^2 + \frac{\varepsilon_0}{2} c^2 B_z^2 - \frac{\varepsilon_0}{2} E_x^2 - n_{e0} m_e v_g \gamma_g \gamma_e v_{ex} - e n_{e0} \gamma_g \phi] = 0\}\bigg|_{PCM} \quad (10b)$$

where we have also made use of Eq. (9), $\mathbf{E} = -\nabla \phi$ and the facts that $v_{ix}|_{PCM} \simeq -v_g$, $n_i|_{PCM} \simeq \gamma_g n_{e0}$ due to the heavy ion mass. The second term in the right hand side of Eq. (7b) is neglected in arriving at Eq. (10b) according to arguments given below equation (7b). Equation (10a) can be further simplified by noting the slow variations of the pulse in the PCM frame. Since we have $E_y = -\partial A_y / \partial t$ and $B_z = -\partial A_y / \partial x$, these quantities become diminishingly small in the PCM frame due to the strong Doppler red shift. This causes the wave momentum density to become negligible compared to the particle momentum density in Eq. (10a). Also the same holds about the time derivative of the wave energy density against the space derivative of electron momentum density. We will return to this problem at the end of this section and for now we take them to be valid. By using above arguments Eq. (10a) is reduce into a constant,

$$[\gamma_e m_e c^2 - e\phi]\big|_{PCM} = \gamma_g m_e c^2. \tag{11a}$$

This is exactly the energy conservation law for particles in the PCM frame [45]. Eliminating $\phi$ in Eq. (10b) by using Eq. (11a) and applying Lorentz transformations $E_x = E_x\big|_{PCM}$ and $\gamma_e = \gamma_e \gamma_g (1 + v_g v_{ex} / c^2)\big|_{PCM}$ in the obtained result, equation (10b) reads as,

$$\{\frac{\partial g_x}{\partial t} = -\frac{\partial}{\partial x}[U_p - C_W]\}\big|_{PCM} \tag{11b}$$

Where $U_p\big|_{PCM} = \varepsilon_0 [E_y^2 + c^2 B_z^2]/2 \big|_{PCM}$ and $C_W\big|_{PCM} = \varepsilon_0 E_x^2 / 2 + n_{e0} m_e c^2 \gamma_e$ are respectively the pulse electromagnetic energy and the well-known wake-field energy [53]. As it will be shown at the end of this section, we could not have ignored the electromagnetic energy and momentum in comparison to the wake-

field energy here, despite to the case of momentum equation. For arbitrary pulse shapes, values of $C_W|_{PCM}$ can be exactly obtained at each time instant by solving the Poisson equation (equation (2) for $v=0$) numerically through the plasma using equations (9) and (11a) for determining density profile in terms of $\phi$. This is the wake excitation problem which is extensively discussed in previous literatures [44,52-57]. $C_W|_{PCM}$ remains constant *behind the laser pulse* [53,54], resulting in the relation $\varepsilon_0 E_x^2/2 + n_{e0}m_e c^2 \gamma_e = \varepsilon_0 E_{wake}^2/2 + n_{e0}m_e c^2$, $E_{wake} \equiv E_{x.\max}$ presents the wake-field amplitude. As we are about considering ultra-short laser pulses, the wake is excited when $L_p < \lambda_p^{NL}$ with resonance condition $L_p = \lambda_p^{NL}/2$ [57]. Here we do not aim to go through the details of wake profile. Important information can be extracted from the overall variations of wave momentum and energy which are given by integration of Eqs. (11a) and (11b) over $x$. In this way, let use the electromagnetic field total energy $H_F = \int_{-\infty}^{\infty} dx u_{EM}$ and momentum $\mathbf{K}_F = \int_{-\infty}^{\infty} dx \mathbf{g}$.

Now integrate equation (11b) at time $t$, over the disturbed plasma region between $x_1 = -v_{ph}t$ and $x_2 = 0$ in the PCM frame. Using the identity $\int_{x_1}^{x_2} dx \{(\partial/\partial t^p)[...]\} = (\partial/\partial t^p)\{\int_{x_1}^{x_2} dx[...]\} - v_{ph}[...]|_{x_1}$ and the fact that pulse vanishes at $x = x_1$ and $x = x_2$, after some straightforward mathematics, we obtain:

$$\left.\frac{dK_F}{dt}\right|_{PCM} = -\frac{1}{2}\varepsilon_0 E_w^2 \tag{12}$$

A localized electromagnetic field as whole can be described as a particle with its total energy $H_F$ and momentum $\mathbf{K}_F$ form a four vector [48, 51]. Based on the basics of relativity theory, we can always write these four-vector in the covariant form of $(H_F, c\mathbf{K}_F) = H_0(\gamma_g, \gamma_g \mathbf{v}_g)$ in terms of the electromagnetic group velocity,

$$\mathbf{v}_g = \frac{\mathbf{K}_F}{H_F}. \tag{13}$$

Here $H_0 \equiv H_F|_{PCM}$ is the electromagnetic energy in PCM frame. Since $H_0$ plays the role of the particle rest mass it remains absolutely constant, i.e. we identically have $\partial H_F / \partial t|_{PCM} \equiv 0$, $H_0$ equals the initial pulse energy in PCM frame. In the same way, the total force exerted on radiation can be written in the covariant form $(f^0, c\mathbf{f}) = (\gamma_g \mathbf{v}_g \cdot \mathbf{F_0}, \gamma_g c\mathbf{F_0})$ [48], where $\mathbf{F_0}$ is the force measured in the PCM frame, i.e. the right hand side of Eq. (12). Using the above arguments equation (12) is transformed into the laboratory frame as, $d[\gamma_g \beta_g]/dt = -\varepsilon_0 \gamma_g c E_w^2 / 2H_0$, with $\beta_g$ being defined as $\beta_g = v_g/c$. Using the identity $d[\gamma_g \beta_g]/dt = \gamma_g^3 d\beta_g/dt$ and the fact that $U_{p0} = \gamma_{g0} H_0$ ($\gamma_{g0} \equiv \gamma_g(t=0)$ $U_{P0} \equiv U_P(t=0)$), this equation takes the form,

$$\frac{d\beta_g}{dt} = -\frac{1}{2} \frac{\varepsilon_0 c \gamma_{g0} E_w^2}{\gamma_g^2 U_{P0}} \tag{14}$$

The above equation, derived for the first time in this work, fully describes the overall evolution of an ultra short laser pulse in the rarefied plasma. It states that pulse is slowly (since $\gamma_g \gg 1$) deaccelerated and loses its energy due to the wake excitation. At small times, when the wake amplitude has not been yet changed noticeably; the solution of Eq. (14) is $\beta_g = \tanh[-t/\tau_d + \alpha]$ where $\tau_d \equiv 2U_{P0}/\varepsilon_0 c \gamma_{g0} E_w^2$ and $\alpha \equiv \tanh^{-1}[\beta_{g0}]$. The full properties of this solution together with its applications to LWFA will be discussed elsewhere. The deacceleration is accompanied by the radiation red-shift. Since the pulse shape does not undergo essential change, its spectral density remains localized around a carrier time-dependent frequency (wave-number) $\omega_0(t)$ ($k_0(t)$) during its propagation. According to the observer in the (PCM) frame, the pulse momentum becomes zero, hence the carrier wave-number should become zero accordingly. In the same way as argued for electromagnetic total energy momentum we can prove that the same covariant four vector presentation for frequency and wave-number is obtained, i.e. we have $(\omega_0, \mathbf{k}_0) = \gamma_g \Omega(1, \mathbf{v}_g/c^2)$ where $\Omega$ is a constant. To convince yourself, you may see that the well-known relation $v_g = c^2/v_\phi$ holds with $v_\phi = \omega_0 / k_0$ being the electromagnetic phase velocity. Therefore we have $\omega_0 = \gamma_g \Omega$. Putting this equation together with the same relation obtained previously for electromagnetic energy $H_F = \gamma_g H_0$ and dividing these two equations we obtain,

$$\frac{H_F(t)}{\omega_0(t)} = \text{constant} \tag{15}$$

which is the adiabatic equation obtained via slow envelop analysis in ref [38]. In this case, the pulse depletion can be completely described by the slow frequency red-shift due to the gradual decrease in the group velocity. The evidences of this theory are given in Fig. 1 which is resulted from a PIC run instance for $n_e / n_c = 0.01$ and $\tau_L = 80 fs$. The wake excitation is shown in Fig. 1(a) where the longitudinal phase-space of electrons together with profile of electric potential are depicted. In Fig. 1(b), the spatial laser pulse profile is depicted in two different times. The horizontal axis is chosen in a way that the pulse front locations coincide for both time instances. On this figure, two arrows are set to point to different locations of a definite pulse hill at different time instances. The increase in the wave-length and the red-shift is clearly observed. In Fig. 1(c), the spectral density of the radiation is plotted for two different time-instances chosen as Fig. 1(b). The total red-shift is shown here by notifying the displacement of central frequency with two pointing arrows at two different times. The physics of this phenomenon can be easily understood; the electrons which are pushed forward by ponderomotive force exert the back reaction to radiation by pushing the field lines backward. This evolution is not a type of light scattering of plasma but rather a depletion. It should be noticed again that Eqs. (14) and (15) describe the average pulse evolutions. Also the defined group velocity $\beta_g$ (Eq. (13) is an average quantity. Trivially, the local variation of electromagnetic energy-momentum and whence the group velocity can deviate from the average trend, but average deviation is zero. An example of this deviation can be observed in Fig 1(b) in the pulse front (see also ref [39]). As will be discussed in section V, this deviation becomes important when steep density gradients are encountered in the case of high density plasma and leads to the light scattering rather than the depletion.

Now we return to discussion of the validity assumptions negligibility of pulse momentum made in simplification of Eq. (10a). The statement can be argued by noting the Lorentz transformations for electromagnetic fields. Using this transformations we have $E_y|_{PCM} = \gamma_g(E_y - \beta_g c B_z)$ and $B_z|_{PCM} = \gamma_g(B_z - \beta_g E_y/c)$. Using the plane wave approximation $E_y \simeq cB_z$ together with these transformations we obtain $g_x|_{PCM} \simeq \varepsilon_0 E_y^2 (1-\beta_g)/c(1+\beta_g)$. The relation between amplitudes of the transverse electric field $E_{y0}$ and the vector potentials $A_0 = A_{y0}$ in terms of the pulse central frequency $\omega_0$ is given by $E_{y0} = \omega_0 A_0$. $A_0$ can be written in terms of its normalized value $a_0 = eA_0 / m_e c$. Using these relations, after the expansion of $\beta_g = (1-\gamma_g^{-2})^{1/2}$ for $\gamma_g \gg 1$ ($\beta_g \simeq 1 - \gamma_g^{-2}/2$), we obtain for the amplitude of $g_x$, say $g_{x0}$, in the PCM frame $g_{x0}|_{PCM} \simeq \varepsilon_0 m_e^2 \omega_0^2 a_0^2 c / 4\gamma_g^2 e^2 = n_{e0} m_e c \omega_0^2 a_0^2 / 4\gamma_g^2 \omega_p^2$. Now we should compare this quantity with the electrons momentum density $m_e \gamma_e n_e v_{ex}|_{PCM}$. Using continuity equation (9) together with the Lorentz transformation for $\gamma_e|_{PCM} = \gamma_g \gamma_e (1 - \beta_g v_{ex}/c) \simeq \gamma_g (c - v_{ex})^{1/2} / (c + v_{ex})^{1/2}$ ($\beta_g \simeq 1$) this quantity reads as $m_e \gamma_e n_e v_{ex}|_{PCM} \simeq n_{e0} m_e c \gamma_g^2 (c - v_{ex})^{1/2} / (c + v_{ex})^{1/2}$. With these arguments, for $g_{x0}|_{PCM}$ to be negligible with respect to $m_e \gamma_e n_e v_{ex}|_{PCM}$ we need to have,

$$\gamma_g^2 \gg \left(\frac{\omega_0^2 a_0}{\gamma_g^2 \omega_p^2}\right) \frac{a_0}{4\chi} \tag{16a}$$

where $\chi$ is defined as $\chi \equiv (c - v_{ex})^{1/2} / (c + v_{ex})^{1/2}$. When this inequality holds, using equation (11a) together with Lorentz transformation $\gamma_e|_{PCM} = \gamma_g \gamma_e (1 - \beta_g \beta_e)$ and approximation $\beta_g \simeq 1$, we obtain $\chi \simeq 1 + e\phi/m_e c^2$. At non-relativistic intensities ($a_0 \ll 1$) the inequality is trivially hold for very rarefied plasmas. At relativistic intensities the factor inside the parenthesis is of order of unity because in this regime the gamma factor of group velocity is approximately given by $\gamma_g \simeq \gamma_L^{1/2} \omega_0 / \omega_p \simeq a_0^{1/2} \omega_0 / \omega_p$. To simplify (16a), let evaluate minimum of $\chi$ which corresponds to the maxima of longitudinal electron velocity. The maximum attainable longitudinal electron velocity in the electron wave is $\beta_g \simeq 1 - \gamma_g^{-2}/2$ which occurs at the *wave break threshold* [53-55]. Using this relation in $\chi$ we find that $\gamma_g \chi > 1/\sqrt{2}$. In this way, the inequality (16a) is assured at relativistic intensities if we have $\gamma_g \gg 1$. Using the estimate $\gamma_g \simeq \gamma_L^{1/2} \omega_0 / \omega_p$ and $\omega_0 / \omega_p = \sqrt{n_c / n_{e0}}$ this inequality takes the form,

$$\frac{n_{e0}}{\gamma_L n_c} \ll 1 \tag{16b}$$

Therefore, because the inequality (16b) holds for very rarefied plasmas, the wave momentum in (10a) can be safely ignored.

In the same way, we can argue that we could not have ignored the electromagnetic energy and momentum in equation (11b). To understand this it should be noted that wake-field energy $C_W|_{PCM}$ is of order of $n_{e0} m_e c^2 \gamma_e$ and the electromagnetic energy $U_p|_{PCM}$ is of the order of $c g_x|_{PCM}$ which is estimated in the last paragraph as $g_{x0}|_{PCM} \simeq n_{e0} m_e c \omega_0^2 a_0^2 / 4 \gamma_g^2 \omega_p^2$. Using this estimate together with the relation $\gamma_g \simeq a_0^{1/2} \omega_0/\omega_p$, we find $U_p|_{PCM} \sim n_{e0} m_e c^2 a_0/8$ which has the same order of magnitude as $C_W|_{PCM}$.

In Fig (1b) it is observed that pulse front undergoes variations other than adiabatic lengthening. The radiation is accumulated in this region. This behavior is due to the local violence of (16b) which causes the unsteady pulse propagation. This phenomenon will be described in details in the section (V) where high density plasmas are concerned.

### B. DISCUSSION OF LONG PULSE BEHAVIOR

In the case of long pulse lengths $L_P > \lambda_p^{NL}$ the pulse is able to excite the high amplitude electron plasma wave (wake-field) as well as the ultra-short pulse. In this case the resonance condition for wake excitation takes the form $L_p = (l+1/2)\lambda_p^{NL}$ [57]. Also, as long as the condition (16b) holds, the pulse depletion is mainly due to the wake excitation and is given by Eqs. (14) and (15). The difference is that when the pulse length exceeds the plasma wavelength the light scattering of a plasma wave, say the Raman type scattering, can make sense. But, due to the very high amplitude of the electron plasma wave (the wake) the light scattering of this wave reveals more complex features compared to the common Raman Scattering (RS). The common RS occurs in the weakly nonlinear (quasi-linear) regime, when density perturbations are very small. In the quasi-linear theory it is the main fundamental mechanism which described light evolutions as long as laser pulse length is short enough to avoid ion response [40-43]. In RS different electromagnetic (EM) modes are coupled to each other via the space charge mode (the electron plasma oscillations), causing energy exchange between the EM modes. In this way, a high amplitude EM mode can feed energy into new forward or backward propagating EM modes causing their growth from the noise level. Beating among the original and new EM modes produces ponderomotive force which resonantly derives plasma oscillations and amplifies the space charge. This causes further amplification of new EM modes and provides the positive feedback to the space charge. In this way, the light propagation through the plasma goes unstable by feeding energy into definite EM and space charge modes and growing them from the noise level. In the case of ultra-intense laser pulses, the electron plasma wave has a much more complex role due to its high amplitude. Note that, here, this mode does not grow from the noise level by means of mode coupling but by the strong laser ponderomotive force. But, nevertheless, the self-modification of the propagating light when penetrating the plasma is expected as well in this case. Because the high amplitude density oscillations greatly alter the plasma dispersion, the group velocity of light becomes a completely local quantity and causes the radiation to undergo self-modulation. The main distinguishing feature of the fully nonlinear regime is that the modulation envelope does not mimic the space-charge pattern. This phenomenon is caused by both the complex shape of relativistic electron wave profile and the locality of group velocity. Concerning the latter, we remind the fact that based on Eq. (14) the group velocity decreases in the pulse front due to seeding of the wakefield. Since the phase velocity of produced wake is equal to the pulse group velocity in this location, the back part of radiation flows forward through the plasma wave. In this way, when a density peak is encountered, the light iss accumulated on its back and is dispersed on its front. PIC simulations at very low densities confirm these statements. As an example we have depicted simulation results for $n_{e0}/n_c = 0.01$, $a_0 = 6$ and $\tau_L = 140 fs$ in Fig. 2. In Fig. (2a).the resultant modulation is observed to have characters described above. Accordingly, the coupling introduced by the space charge between different electromagnetic modes is not of resonant type. This causes not a few but an infinite number of EM modes in the light spectrum to become coupled to each other, i.e. the frequencies of scattered EM modes does not obey the common relation $\omega_0 \pm \omega_p$ but rather by the relation $\omega_0 \pm \ell \omega_p^{NL}$ with $\ell = 1, 2...$ and (see also [58]) $\omega_p^{NL} \simeq \omega_p / \sqrt{\langle \gamma_L \rangle}$ [46, 59], $\langle \gamma_L \rangle \simeq \sqrt{1 + a_0^2/2}$. The factor $\langle \gamma_L \rangle$ is introduced in plasma frequency to account for relativistic mass increase. The nonlinear plasma frequency in this case is $\omega_p^{NL} \simeq 0.48 \omega_p$. Using this calculation, the growth of modes with frequencies $\omega_0 \pm \ell \omega_p^{NL}$ is clearly observed in Fig. (2b) where we have depicted the spectral density of radiation. The amplified modes are specified by arrow on the figure. Because the intercalation between the scattered light and plasma wave is of non-resonant type, this phenomenon does not result in instability but instead very slow pulse modulational. When the pulse length is not large enough, the accumulation of scattered light does not introduce significant radiation back-reaction on the plasma wave. When very long pulse lengths are considered, the radiation back reaction becomes noticeable and causes induced wave breaking, say the radiation pressure induced wave breaking (RPIWB). At these conditions, the scattering evolves from Raman type to Thomson type. The stimulation mechanism is related to the disintegrability of the electron motion in the presence of back-scattered light which is extensively discussed in recent literatures [33-35]. The difference between very long and long pulse behaviors has been described in detail elsewhere [60]. When the plasma density grows toward critical value the plasma optical response becomes strong enough to cause much faster RPIWB, causing its occurrence with short pulse lasers. This is described in the next section.

## V. NEAR CRITICAL PLASMA DENSITIES

### A. NONLINEAR ENERGY DEPLETION AND ANOMALOUS LIGHT SCATTERING

From arguments given in section IV.A, if the condition of the Eq. (16b) becomes violated, say when $n_{e0}/\gamma_L n_c$ grows toward unity, the adiabatic pulse depletion becomes doubtful. At these conditions, the pulse can undergo rapid changes during its propagation. For parameters of current intense lasers, high densities correspond to the regime of $L_P > \lambda_p^{NL}$. In section IV.B., it has been noticed that the light scattering off the plasma wave behaves strangely when high amplitude waves are encountered. At very low densities, the light scattering does not introduce noticeable modification in the space charge mode due to the very slow development. In spite, when the light scattering becomes rapid due to the rise in $n_{e0}/\gamma_L n_c$, serious modifications are encountered. Here, those beating between scattered EM modes which are non-resonant with the space charge can result in strong wave breaking. When a density peak is encountered, the light is accumulated on its back and is dispersed on its front as already we have described on Fig. (2a). As a result, a net forward radiation pressure is exerted on the density column. This pressure accelerate the plasma electrons in the direction of the plasma wave phase velocity [58]. Under these conditions, the electron plasma wave undergoes early wave breaking as observed in Ref. [61]. When the pulse length, $L_p$, exceeds noticeably the plasma wave length $\lambda_p = c/\omega_p$, the wave break strongly affects the plasma oscillations causing them to undergo wild modulation.

This leads to inapplicability of the concept of radiation scattering off a plasma wave. Naturally, at these conditions, the phase-matching relations between the radiation and space charge disappear. Furthermore, wave breaking results in the saturation of the plasma scattering and the plasma back reflection [62]. In the pulse front electrons are accumulated by the laser ponderomotive force resulting in a positive density gradient. This density gradient can accumulate and even back reflect the light. The distinctive feature of the front is that the wave-breaking is absent because of no electrostatic field being exist beyond the front. Thus the plasma scattering cannot be saturated by wave-breaking. In this location, the plasma can reflect the light in the same way as it does in the radiation trap inside the density valleys described in Ref. [63]. If the peak density in the frame co-moving with it, reaches the critical value $n_c$ in this frame, the reflection takes place. Since the positive density gradients are amplified by receiving positive feedback from radiation, the reflection eventually occurs in the absence of dissipative processes. However, the laser energy conversion to the electrostatic and kinetic energy of electrons exists as a dissipating process as discussed in IV.A. When the density scale length is short enough, the accumulation can gain over the dissipation and eventually can lead to the back reflection. The density scale length can be estimated by the nonlinear plasma wavelength $\lambda_p^{NL} = Fc/\omega_{pe}$. The time needed for reflection is given by $\tau_r \sim \lambda_p^{NL}/2\Delta v_g$, with $\Delta v_g$ being the deviation of the maximum local group velocity with respect to the average value given by Eq. (13). Note that the maximum of local group velocity occurs at the density minimum displaced by $\lambda_p^{NL}/2$ with respect to the front. When the reflection time is shorter than the depletion time, $\tau_d$, the pulse reflection takes place. And this occurs again when the $n_{e0}/\gamma_L n_c$ grows toward unity. Our PIC simulations mentioned in Sec. II show when the plasma density is raised toward values of order of $0.1 n_c$ and higher, the electron pile-up formed in the pulse front, quickly becomes reflective and then propagates as a solitary disturbance along the plasma. The longitudinal electric field produced by this pile-up converts the laser energy into the wave break electrons continuously. This provides an indirect mechanism for electron acceleration. The formation and propagation of the electron soliton (pile-up) can be directly observed in PIC simulation results presented in Fig. 3. In panels (a) and (c) the longitudinal phase space together with vector potential and electron density profile are sketched for $a_0 = 6$, $n_e/n_c = 0.3$ and $\tau_L = 100\, fs$ at two different time instances. In panels (b) and (d) the longitudinal electric field is plotted together with the density profile in the location of electron pile-up, showing the double-layer structure of the perturbation. Electron acceleration by the formed structure is also directly observed in this figure. The radiation becomes gradually reflected by the electron pile-up. In the soliton rest (SR) frame, the radiation is the superposition of the approximately equal-amplitude incident and reflected lights. The electron pile-up is conserved by the radiation pressure produced by the light reflection.

Let the soliton to propagate with velocity $v_s$. The incident pulse in the SR frame experiences the relativistic frequency down-shift with respect to the lab frame, given by $\omega_i|_{SR} = \omega_0 (c-v_s)^{1/2}/(c+v_s)^{1/2}$. Hereafter subscripts $i$ and $r$ are used to present incident and reflected quantities, respectively. The notation $X|_{SR}$ is used to indicate that $X$ is measured in the SR frame. The pulse length in the SR frame can be obtained using the invariant normalized phase number, $N_\theta \equiv \Delta\theta/2\pi$, where $\Delta\theta$ is the phase difference between the start and the end of the pulse. In terms of this quantity, we have $L_{p,i}|_{SR} = N_\theta \lambda_i |_{SR} = L_p \omega_0/(\omega_i)|_{SR} = L_p (c+v_s)^{1/2}/(c-v_s)^{1/2}$. In transformation from SR into the lab frame, the reflected radiation experiences another frequency down-shifted due to the negative relative frame velocity $-v_s$, i.e. $\omega_r = \omega_i|_{SR}(c-v_s)^{1/2}/(c+v_s)^{1/2}$. Therefore, the back reflection in the lab frame receives the double frequency down-shifted,

$$\omega_r = \omega_0 \frac{(c-v_s)}{(c+v_s)}. \tag{17}$$

Correspondingly, in the same way as discussed above for the incident pulse, the reflected pulse becomes stretched with respect to the initial pulse as follow,

$$L_{p.r} = L_p \frac{(c+v_s)}{(c-v_s)}. \tag{18}$$

Since the transverse vector potential remains invariant in Lorentz transformations, the expression $E_y/\omega_0$ also should remain invariant. Hence the electromagnetic energy density of the reflected radiation, $u_{EM.r}$, is reduced by a factor of $(\omega_r/\omega_0)^2$ with respect to the incident radiation. Using this argument with together equations (17) and (18), the *reflectivity* ($R$), the fraction of the total reflected electromagnetic energy ($H_{F.r} \sim u_{EM.r} L_{p.r}$) over the incident energy ($H_{F.i} \sim u_{em.r} L_{p.i}$) is estimated to be:

$$R \sim \frac{c-v_s}{c+v_s}. \tag{19a}$$

The *absorption coefficient*, $\eta = 1-R$, is estimated to be,

$$\eta \sim \frac{2v_s}{c+v_s}. \tag{19b}$$

Note that stored electromagnetic energy in terms of longitudinal electric field has been ignored in above derivation. As will be shown in simulation results, this ignorance is satisfactory.

In the physical model presented so far, the interaction ends up as soon as the pulse is being completely reflected, i.e. an absorption saturation is attained by the reflection completion. The saturation time, $t_{sat}$, can be obtained easily by calculating the time-difference between two relativistic events (time-space points) in the laboratory frame which correspond to the begin and end of the pulse reflection in the SR frame. These events occur respectively at $(t_1, x_1)|_{SR} = (0,0)$ and $(t_2, x_2)|_{SR} = (L_{p.i}/c, 0)|_{SR}$. Time difference between these events in the laboratory frame is given by Lorentz time transformation, $t_{sat} = \gamma_s L_{p.i}|_{SR}/c$, with $\gamma_s$ being $\gamma_s = (1 - v_s^2/c^2)^{-1/2}$. Therefore, using relation for $L_{p.i}^s$ given in the last paragraph we have,

$$t_{sat} = \frac{L_p}{c(1 - v_s/c)}. \tag{19c}$$

An estimate of plasma temperature in terms of physical parameters $n_{e0}$ and $a_0$ can be made using obtained relations for $\eta$ and $t_{sat}$ in the form of $k_B T_{e.h} \sim \eta H_{F0}/n_{e0} v_s t_{sat}$, with $H_{F0}$ being the initial electromagnetic energy. In this way, using equations (19), this estimate reads as,

$$k_B T_e \sim 2\frac{\langle u_{em0} \rangle}{n_{e0}} \frac{c - v_s}{c + v_s} = \kappa m_e c^2 \left(\frac{\omega_0}{\omega_p} a_0\right)^2 \frac{c - v_s}{c + v_s} \tag{20}$$

where $\kappa = \int a^2 dx / a_0^2 L_p$, calculated at $t=0$, is the pulse shape factor ($\kappa \leq 1$, $\kappa = 1$ for rectangular shape pulse).

We can also made an estimate of the soliton velocity, $v_s$. Since the electromagnetic energy remains conserved in the SR frame, according to discussions given in section IV.A above Eq. (14), this frame is the special frame of radiation which moves with the group velocity of Eq. (13), i.e. $v_s = v_g$. By assuming local equilibrium between plasma and radiation inside the first plasma cavity behind the electron pileup, the velocity of electron soliton can be estimated using equations (10a) and (10b). The local equilibrium means $\partial \gamma_e n_e / \partial t |_{PCM} = 0$, $\partial \gamma_e n_e v_{ex} / \partial t |_{PCM} = 0$, $\partial u_{EM} / \partial t |_{PCM} = 0$ and $\partial g_x / \partial t |_{PCM} = 0$. In the other hand, since the electromagnetic field in the PCM frame is a standing wave, formed by superposition of incident and reflected wave, its momentum is zero in average. In this way equation (10a) reduces to the wake constant (11a) again. Using this statement, the problem of obtaining induced density profile reduces to solving for wake excitation by a pulse with its amplitude (the standing wave amplitude), $a_0'$, being two times the original laser amplitude ($a_0' = 2a_0$). Hereafter the prime is used for the standing wave quantities. Now using equation (11b) and arguments given below this equation we obtain in the plasma region considered here

$U_p'\big|_{PCM} = \varepsilon_0 E_{wake}^2 / 2$. Recalling arguments given above equation (16a), we have

$U_p'\big|_{PCM} \approx \varepsilon_0 E_y'^2 (1-\beta_g)/(1+\beta_g) \approx \varepsilon_0 \omega_0^2 m_e^2 c^2 a_0'^2 / 4e^2 \gamma_g^2$. On the other hand we have

$E_{wake} = \sqrt{2(\gamma_{e.max}-1)} m_e c \omega_p / e$ [56, 57] where $\gamma_{e.max}$ is the maximum electron gamma factor inside the wake. Using these last two equations into the former we obtain,

$$\gamma_g \approx \frac{a_0}{\sqrt{\gamma_{e.max}-1}} \frac{\omega_0}{\omega_p} \qquad (21)$$

At weak laser intensities ($a_0 \ll 1$), $\gamma_{e.max} \approx 1 + a_0^2$ and we recover the well known linear result. At ultra relativistic intensities ($a_0 \gg 1$), $\gamma_{e.max}$ scales as $\gamma_{e.max} \sim a_0$ and Eq. (21) scales as the well known relation $\gamma_g \sim a_0^{1/2} \omega_0 / \omega_p$. This equation shows increase against the laser intensity and decrease against the plasma density. As will be shown this trends are confirmed by our simulations.

The important features of the radiation dynamic are visualized in Figs. 4 and 5 using PIC simulations. In panels (a) and (b) the spatiotemporal evolution of the vector potential is presented for parameters same as Fig. 3. The radiation gradual back reflection and overall de-acceleration and lengthening of the reflected pulse can be easily seen in these plots. In panel (c) the history of the vicinity of the radiation front is sketched for two values of intensity $a_0 = 6, 10$ and two values of density $n_e / n_c = 0.3, 0.5$, with other parameters being chosen as (a) and (b). This plot verifies the predictions of (21) about decrease (increase) of group velocity versus increasing the plasma density (laser intensity). In Fig. 5, the spatiotemporal evolutions of the spectral density of the vector potential are presented for parameters same as Fig. 5a. The figure shows the gradual depopulation at the original laser wav-number $k = k_0$ and population at lower wave numbers. This is consistent with the model of light reflection from the moving radiation front. The broadening of the radiation spectrum bellow $k = k_0$ is due to complex motion of electrons in the strong radiation which causes strong harmonic generation [57]. The saturation of reflection is also seen in this plot as the spectrum shape remains invariant after a certain time. In Fig. 6 we described different aspects of the radiation absorption. In Figs. 6(a) and 6(b), we have shown time histories for electromagnetic and plasma energies and momentums for $a_0 = 6$, $n_{e0}/n_c = 0.3$ and $\tau_L = 100 fs$. Both the total energy-momentum conservation and absorption saturation are seen in this figures. In Fig. 6(b), it is shown that total energy inside the box is decreased due to the back-reflection. It is also shown that the stored energy in the longitudinal part of electric field is neglectable compared to the absorbed energy, an assumption which has been made in the proposed model.

The validity of the proposed model can be best demonstrated by changing the pulse length and keeping the other parameters constant. Since $v_s$ depends only on the plasma density and pulse intensity, the absorption should remain constant as is predicted by (19b) and the saturation time should change exactly proportional to the pulse length as is predicted by (19c). In Fig. 6(c), the plasma energy (absorption amount) is plotted at the laser intensity $a_0 = 6$ for two plasma densities, $n_{e0}/n_c = 0.3$ and $n_{e0}/n_c = 0.5$, and two pulse lengths $\tau_L = 100 fs$ and $\tau_L = 200 fs$. The saturation times are pointed by arrows in this figure with their values being presented. On can see that presented values are in excellent agreement with the model formulas (19b) and (19c) in the sense described just above.

At this end let further compare predictions of model equations with direct simulation results. As can be read from Fig. 3(b), the position of the formed electron soliton (pile up) is $x_{s.f} \simeq 107 \mu m$ at time $t = 300 fs$. The initial position at $t = 0$ is the left plasma boundary $x_{s.i} = 40 \mu m$. The average soliton velocity is $\bar{v}_s = (x_{s.f} - x_{s.i})/(t_f - t_i) \simeq 0.74c$. Using this value as an estimate in equation (19b) we get $\eta \approx 0.85$ which is in good agreement with the direct result $\eta \simeq 0.75$. The origin of extra absorption calculated via (19b) is due to the ignorance of radiation back-reflection from plasma parts other than the electron pile-up. A part of the light is also reflected in the process of soliton formation. From equation (19c), the saturation time is estimated to be $\omega t_{sat} \simeq 728$ ($t_{sat} \simeq 386 fs$) which is very close to the direct result, $\omega t_{sat} \simeq 680$ (see Fig. 6(c)). The calculated temperature from equation

(20) is $k_B T_{eh} \simeq \kappa \times 9.2 \text{MeV}$. With $\kappa \simeq 0.7$ which is typical of our pulse shape, we obtain $k_B T_{eh} \simeq 6.4 \text{MeV}$. The direct result is $k_B T_{eh} \simeq 4.5 \text{MeV}$. The error is due to inaccuracy in the calculated absorption discussed above.

At the end of this subsection we wish to add comments on the description of plasma heating in the proposed model. In this model equations (19b) only indirect laser heating mediated by the longitudinal electric field of the formed double layer (see Fig. 3) has been taken into account. Based on group velocity behaviors described by Eq. (21) and Eq. (19b), the overall heating (total absorption) is predicted to decrease with increase in the density. This trend is already confirmed in our simulation results for absorption history given for example in Fig. 6(c) for different plasma densities, $n_{e0}/n_c = 0.3$ and $n_{e0}/n_c = 0.5$. Based on indirect electron acceleration mechanism used in the proposed model, we expect that slope of the absorption-curve has the same behavior. However simulations result does not satisfy this expectation. Quiet reverse, it is seen in Fig. (6c) that this slope is larger at the higher plasma density ($n_{e0}/n_c = 0.5$). This increase in the sloop is in fact due to the direct laser absorption by electrons which has not been taken into account in the Eq. (19b). The direct laser absorption occurs when freed (unbounded) electrons enter the radiation region (see Fig. 3) [29-34]. A more detailed discussion of the plasma heating is given in the next subsection.

## B. PLASMA HEATING AND SUPER-PONDEROMOTIVE ELECTRON GENERATION

The electron acceleration by intense laser inside the under-dense plasma has been extensively studied previously, with the large majority of attentions having been paid to the observed super-ponderomotive electrons (electrons with energies larger that $m_e c^2 a_0^2 / 2$). The energy scale of these electrons is beyond the scale of attainable energy inside the plasma wave body ( of order of $\sim m_e c^2 a_0$) and their total current is very larger than what producible by the wakefield acceleration [3-5, 52]. In this respect, a variety of *direct laser acceleration* (DLA) mechanisms have been proposed for description of these electrons, including acceleration in the ion channel [29-32], stochastic acceleration by evolving radiation [33-35] and enhanced electron acceleration by anti-dephasing effects [37, 64]. Here, our proposed radiation model suggests a new mechanism for plasma heating and super-ponderomotive electron generation. The radiation induced wave breaking, described in IV.B and V.A, plays a primary role in this mechanism. Based on arguments presented in the last subsection, the laser absorption is dominated by the indirect acceleration of wave break electrons in the space charge field sustained by the radiation pressure. The phase mixing attained by intersection between trajectories of the wave break electrons and oscillation body electrons leas to the volumetric plasma heating. Furthermore, a new direct laser acceleration mechanism can be identified based on described radiation model. When the pulse length exceeds the plasma wavelength electrons can be liberated inside the radiation region by means of wave breaking (see Fig. 3). In the PCM frame the radiation is described by a standing wave produced by superposition of equal amplitude incident and reflected lights. It has been shown previously that the electron motion inside the standing wave is disintegrable and can become chaotic at nonlinear regime [35]. These characters of motion can lead to stochastic energy gain by electron [33-34]. Therefore, inside the PCM frame electrons are directly heated by laser via the stochastic mechanism. Inside this frame, the maximum attainable energy can be as large as the ponderomotive scale $m_e c^2 (2a_0)^2 / 2 = 2 m_e c^2 a_0^2$. Since the PCM frame moves with velocity $v_g$ with respect to the laboratory, the results for forwardly directed electrons are boosted by a factor of $\gamma_g$ in the laboratory frame. For example the Lorentz transformations for momentum implies,

$$p_{ex} = \gamma_g \gamma_{e.st} (v_g + v_{ex.st}), \tag{22}$$

with $v_{ex.st}$ and $\gamma_{e.st}$ being respectively the stochastic electron velocity and gamma factor in the PCM frame. This equation predicts that maximum electron energy increases versus $\gamma_g$ for a definite laser intensity, a result which will be confirmed by our simulations (see V.C). Also it predicts larger absolute momentum values for forwardly directed electrons ($v_{ex.st} > 0$) with respect to backwardly directed ones ($v_{ex.st} > 0$). The fact that can be already seen in Fig. 3 and is frequently observed in previous publications. Furthermore it is inferred from Eq. (22) that electron can attain energy of several times the ponderomotive value, as frequently observed previously. The DLA mechanism proposed here may be termed as the *relativistically boosted stochastic heating*.

In Fig. 7, the phase variables of two selected test electrons are presented for the simulation instance of $a_0 = 6$, $n_e/n_c = 0.5$ and $\tau_L = 100 fs$. The $t-x$ and $t-p_y$ diagrams presented in panels (a) and (b) demonstrate the free (non-bounded) motion of test particles in the radiation field. Fig. 7(a) shows the test particle initially located at the left plasma boundary enters the free motion after a transient bounded phase. It approximately travels the whole disturbed region along the pulse propagation. The second particle moves freely opposite to the pulse propagation until it reaches to the plasma boundary. After then, it is returned back by the space charge (see Fig. 7(b)). This particle is a part of the return current which is produced due to the strong wave-breaking. In Figs. 7(c) and 7(d), $p_x - p_y$ phase slice is plotted for test electrons. It is clearly seen that after passing several chaotic phases, the test electron eventually experiences the DLA and gains energies beyond the ponderomotive scale.

There are three types of motion inside the disturbed plasma which correspondingly produce three types of population in the electron spectrum. The first type is the motion inside the electrostatic oscillation-body. The second type belongs to the electrons which are produced in the wave-break and exchange energy with plasma oscillations. The last type corresponds to the most energetic free electrons which undergo the DLA mechanism. In Fig. 8, we have given PIC simulation results for electron spectrum at four different times for $a_0 = 6$, $n_e / n_c = 0.3$ and $\tau_L = 100 fs$. Since the disturbed region is increased in length due to laser propagation, the overall hot electron populations increase accordingly. To understand heating process in the course of time, we refer to the simulation results given in Fig. 9. Here, we have presented time histories of electron temperature and maximum gamma factor for different plasma densities and two laser intensities ($a_0 = 6$ and $a_0 = 10$) at $\tau_L = 100 fs$. It is observed that the electron temperature is approximately saturated at a definite time, say $t'_{sat}$. Afterward it grows very slowly up to the end of the interaction (absorption saturation $t_{sat}$). Our additional simulations performed with different pulse shapes (not given here) have indicated that the shape of temperature history before the saturation time is directly related to the pulse shape in its rising part. Especially $t'_{sat}$ is proportional to the pulse rise time, as is expected according to arguments we have given about Eq. (19c) . The maximum gamma factor shows the same behavior as the temperature, except for fluctuations in the saturation regime.

### C. Simulation Trends

In Fig. 10, we have depicted the total absorbed electromagnetic energy, $\eta$, versus the plasma density (in panels (a) and (b) for $a_0$=6 and $a_0$=10, respectively) and laser intensity (in panel (c) for $n_{e0}/n_c = 0.5$). The corresponding saturation times, $t_{sat}$, are also given in this figure. It is noticed that in agreement with our model Eqs. (19b), (19c), (20) and (21), both the absorbed energy and the saturation time decrease versus the plasma density and increase versus the laser intensity. Note that the electromagnetic group-velocity given by Eq. (21) decreases (increases) by increasing the plasma density (laser intensity).

Trends for saturated value of electron temperature and corresponding saturation time versus the plasma density and the laser intensity are given in Fig. 11. Parameters are chosen as Fig. 10. Again, both quantities decrease versus the plasma density (Fig. 11(a)) and increase versus the laser intensity (Fig. 11(b)). For the same parameters, trends for the maximum electron energy are given in Fig. 12, in panel (a) versus the density and in panel (b) versus the intensity. Observed trends are the same as those of temperature and confirm the model equation (22).

All trends given above indicate that the total absorption is mainly dominated by the saturation time. Larger saturation time means larger absorption. Equally speaking the laser absorption is dominated by the indirect mechanism described in section IV.A.

## VI. CONCLUSION AND DISCUSSION

In conclusion, we have presented a comprehensive one dimensional theory of intense short-pulse laser interaction with under-dense plasma. The kinetic aspects of plasma have been fully incorporated using extensive particle-in-cell (PIC) simulations. As described in the introduction, our study might be very important in understanding of the recent laser plasma experiments. Also, results presented here could find applications in the fundamental theory of intense light propagation through under-dense plasma. Based on this study, it has been turned out that the plasma behavior is characterized by both its density and the ratio of the pulse length to the plasma wavelength. In this respect, different interaction regimes have been identified and described in details in sections IV-V. Accurate estimates for the overall laser absorption and plasma heating have been obtained for different regimes identified in this study. Though our study is one dimensional, the fundamental phenomena described here can occur as well in multi-dimensions. The description of light depletion by wake excitation and light scattering by induced plasma profiles remains independent of space dimensions. In the same way, the important role of these phenomena in plasma heating remains unchanged when multi-dimensional space is considered. The most important evidences for these statements come from very recent experiments carried out using sub-critical foam targets [16, 25, 27]. In these experiments, it has been demonstrated that the penetration depth is the most important determining factor in the total laser absorption and overall plasma heating. And it has been demonstrated that this depth increases with density decrease. These outcomes are both in agreement with our results. The agreement confirms the fact that the multi-dimensional effects cannot rule out the phenomena described here but they can only cause modifications. In multi-dimensions, the laser light can undergoes self-focusing [27]. If this self focusing could be improved by density increasing and this improvement could eventually result in further radiation imprinsment inside the induced plasma, it might be able to modify the simulation trends observed in V.C. Moreover, the direct electron acceleration may become parametrically amplified via betatron oscillation resonance in the multi-dimensions [36-37, 65]. Because these electrons are poorly populated, this phenomenon is unlikely to introduce significant effects. An important aspect of multi-dimension which may seem more relevant to our study is the instability of the critical surface [66]. In the case of laser propagation through under-dense plasma, this instability can lead to multi branching of the propagation [67]. Since our description holds about each radiation branch, the multi branching cannot alter our results seriously. The subject of these multi-dimensional effects is currently under our investigation as a new work.

## ACKNOWLEDGMENT

Authors E. Y., A. C. and E. K. would like to appreciate the leadership of Dr. J. Yazdanpanah in the course of this research.

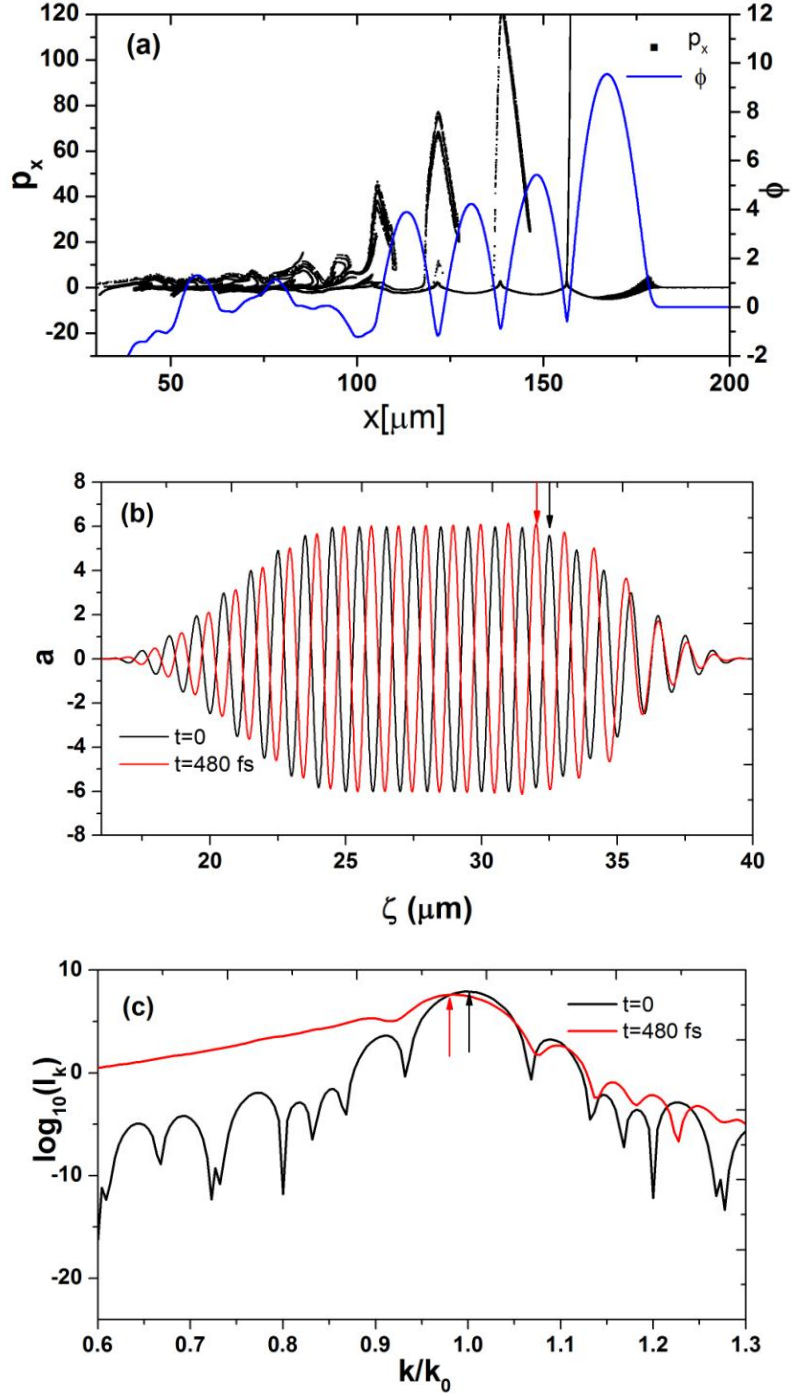

**FIG. 1.** (Color online) Longitudinal phase-space $x - p_x$ (left axis) and potential $\phi$ (right axis) (a), laser pulse at two different time instances (b), and the pulse spectral density at these times (c), all for $a_0 = 6$, $n_{e0}/n_c = 0.01$ and $\tau_L = 80\,fs$. In the panel (b) the horizontal axis shows $\xi = x - v_f t$, with $v_f$ being the pulse-front velocity. Arrows in panels (b) and (c) show different locations of a definite maximum and the carrier frequency respectively.

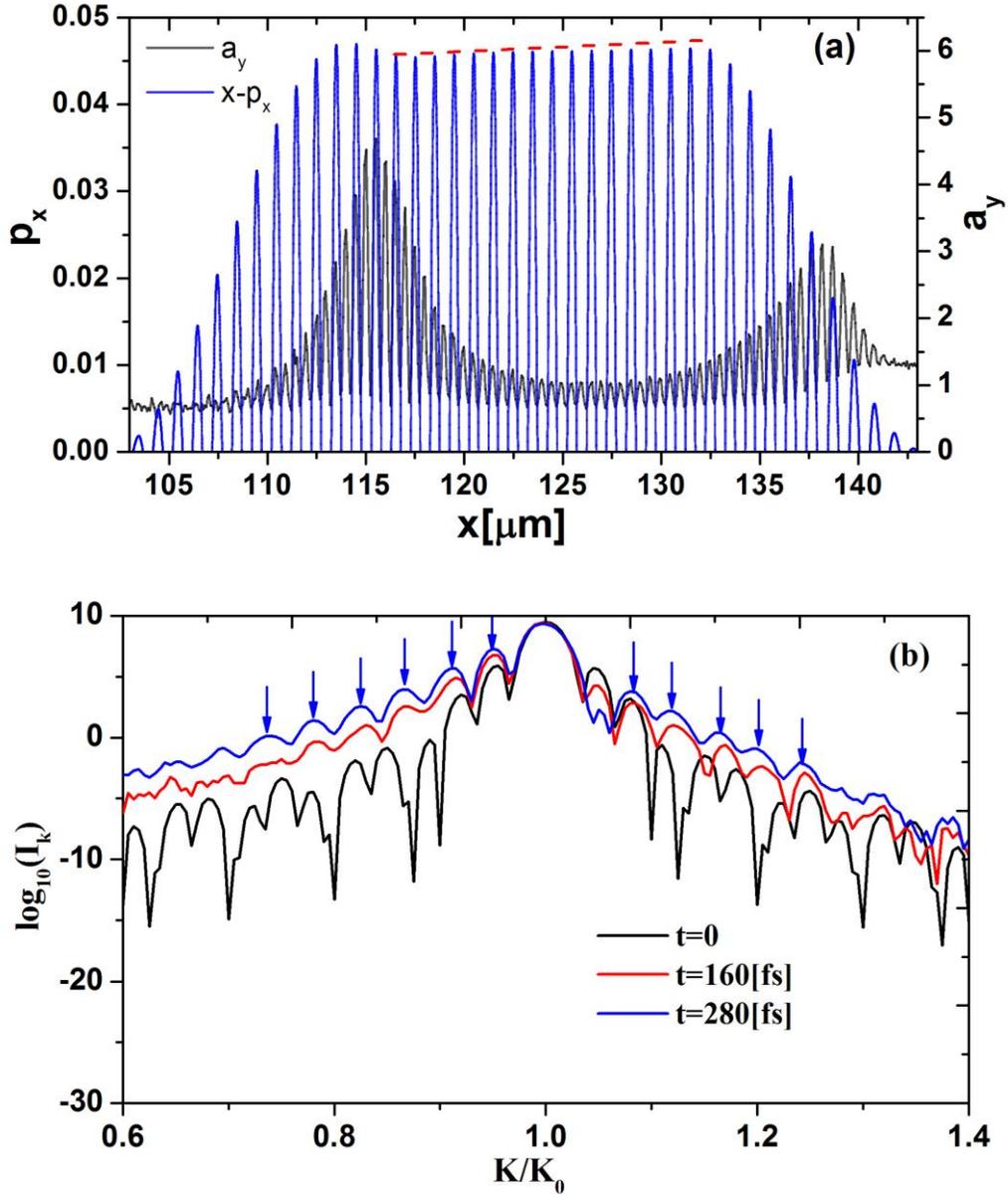

**FIG. 2. (Color online) The pulse profile in the positive half-plan (left axis) together with $x-p_x$ phase space at $t=280\,fs$ (a), and the spectral density of radiation at different times (b), for parameters same as Fig.1 but with $\tau_L=140\,fs$. In (a) the dashed line is for eye-guiding. In (b) locations of scattered modes are marked by arrows.**

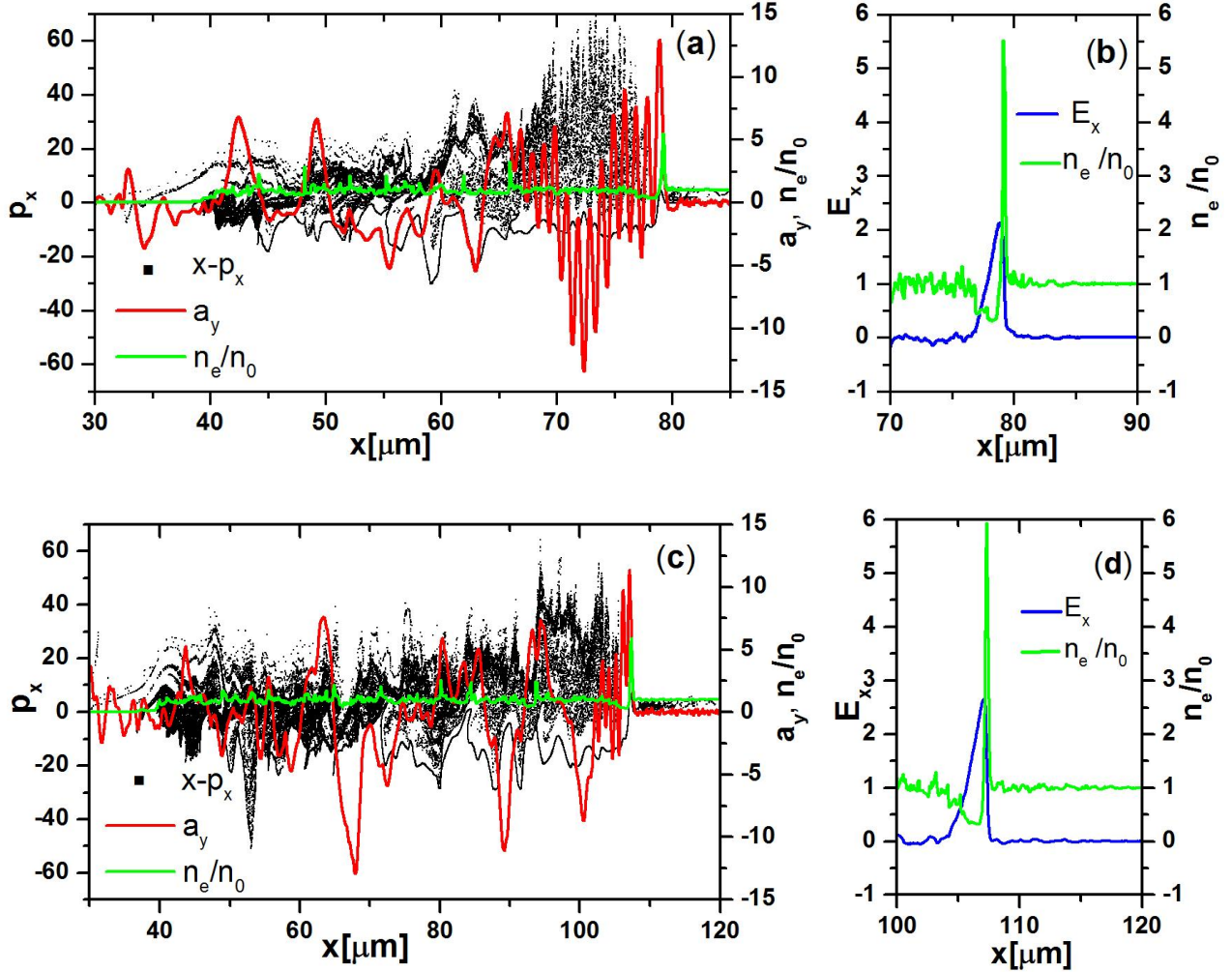

**FIG. 3. (Color online) The longitudinal electron phase space (left axis), the vector potential (right axis) and the electron density profile (right axis) (a), and the longitudinal electric field (left axis) and electron density (right axis) in the front region (b), at $t = 180\,fs$, together with plots for the same issues at $t = 300\,fs$ (c, d). Here $a_0 = 6$, $n_{e0}/n_c = 0.3$ and $\tau_L = 100\,fs$.**

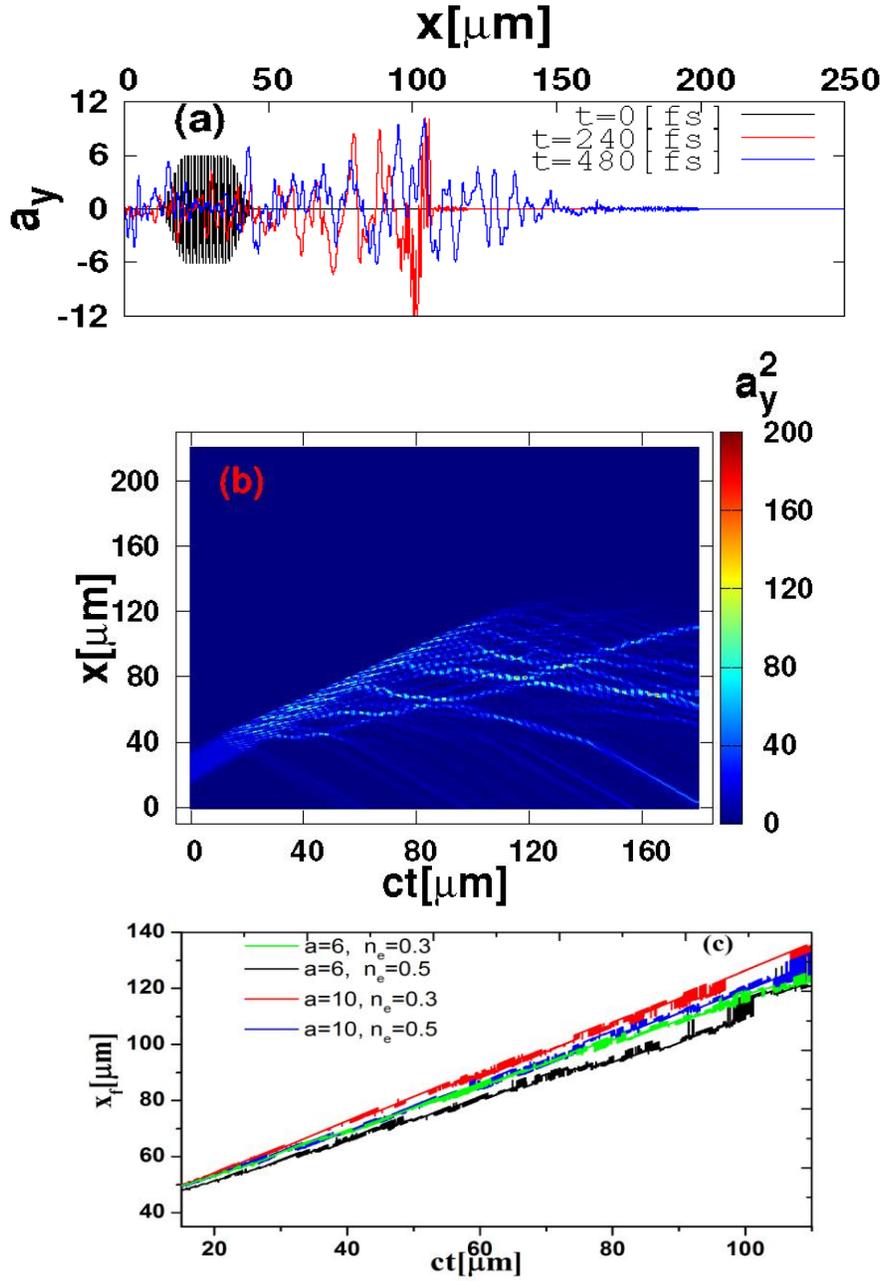

**FIG. 4. (Color online) The spatiotemporal evolutions of electromagnetic wave; spatial profile of the vector potential at different time instants (a), the full histogram of the vector potential intensity over the whole simulation time (b), and the history of radiation front location for different laser intensities and plasma densities. In (a) and (b) parameters are same as Fig. 3. In (c) $\tau_L = 100 fs$.**

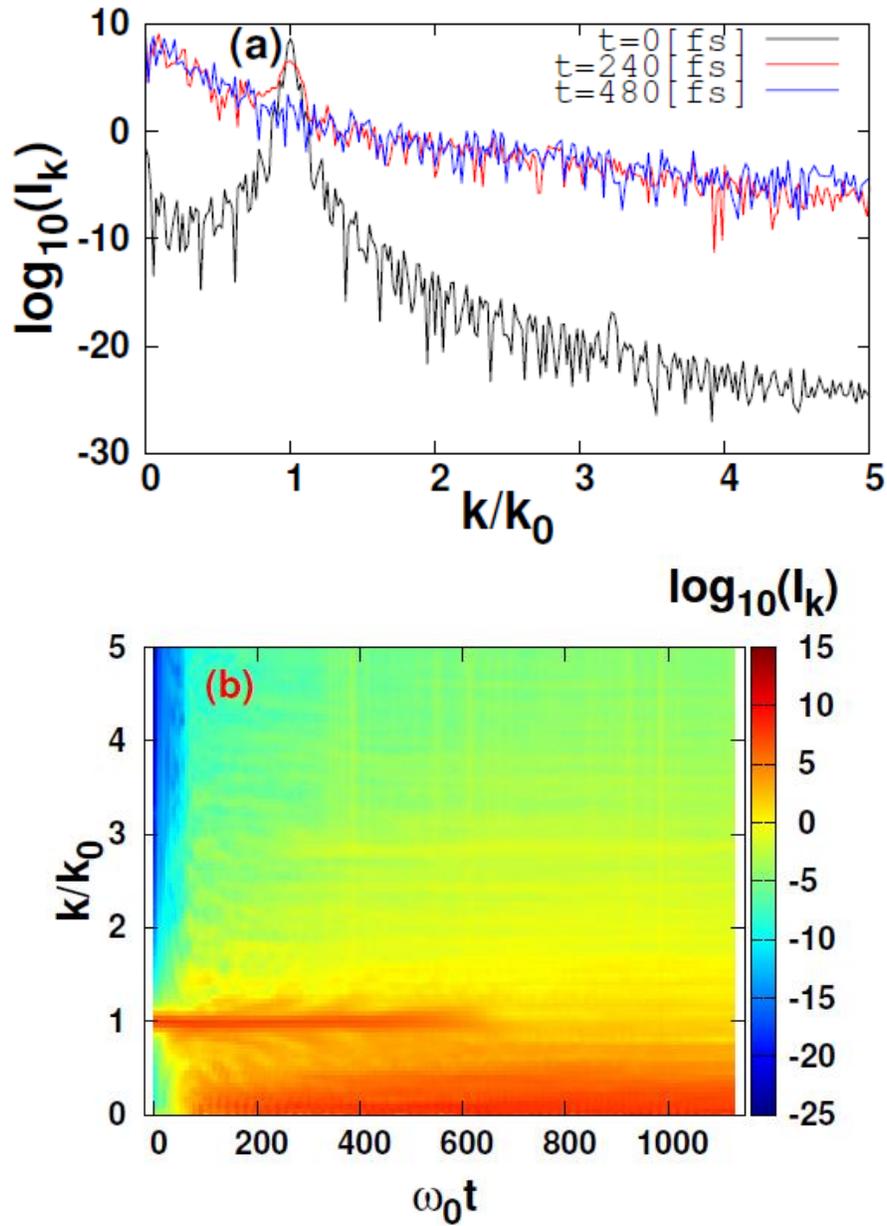

**FIG. 5. (Color online) Evolutions of the radiation spectral density; spectral density at different time instants (a), and its full histogram over the whole simulation time (b), parameters are same as Fig. 3.**

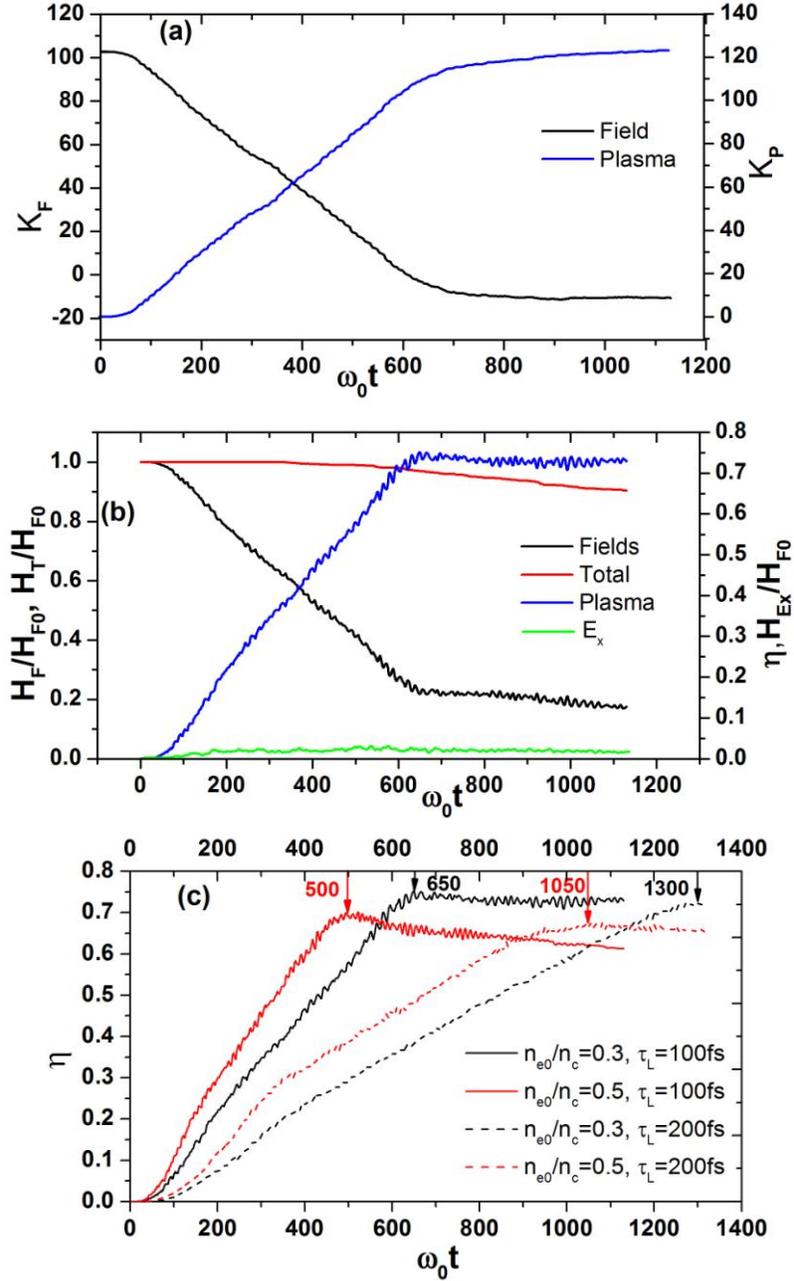

**FIG. 6. (Color online) The histories for field (left axis) and plasma (right axis) momentums (a) and energies (b) both for, $a_0 = 6$, $n_{e0}/n_c = 0.3$ and $\tau_L = 100\,fs$, and histories for plasma energy at the same laser intensity but two different plasma densities $n_{e0}/n_c = 0.3$ and $n_{e0}/n_c = 0.5$ and two different pulse lengths $\tau_L = 100\,fs$ and $\tau_L = 200\,fs$ (c). In panel (b), the total energy inside the simulation box (left axis) and energy of longitudinal electric-field (right axis) are also depicted.**

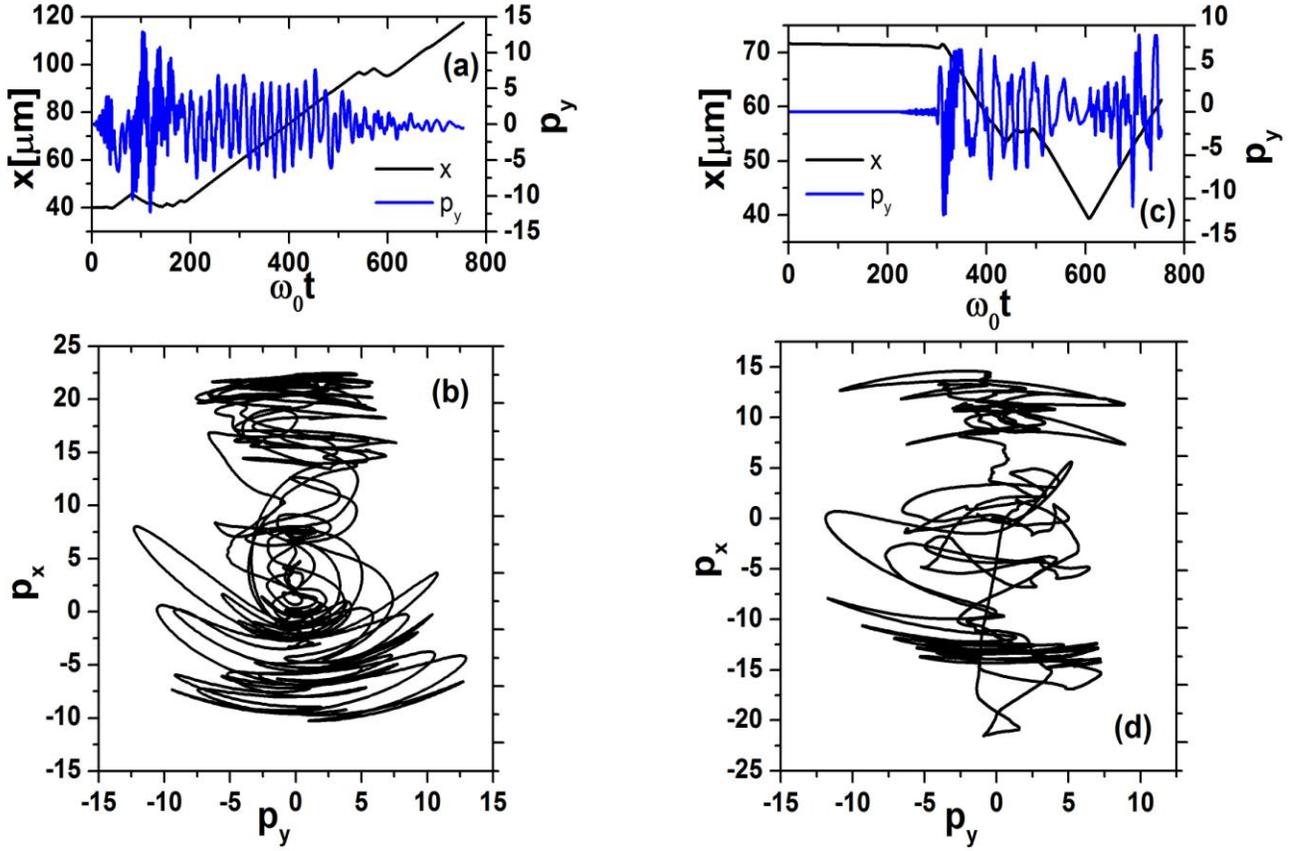

**FIG. 7. (Color online)** $t-x$, $t-p_y$ (a, b) and corresponding $p_x-p_y$ diagrams (c, d) for two selected test particles with laser plasma parameters $a_0=6$, $n_{e0}/n_c=0.5$ and $\tau_L=100\,fs$ and initial conditions $(x_0, y_0, v_{x0}, v_{y0}) = (40, 0, 4.3e-4, 7.3e-3)$ **and** $(x_0, y_0, v_{x0}, v_{y0}) = (71.58, 0, 6.5e-3, 1.6e-3)$.

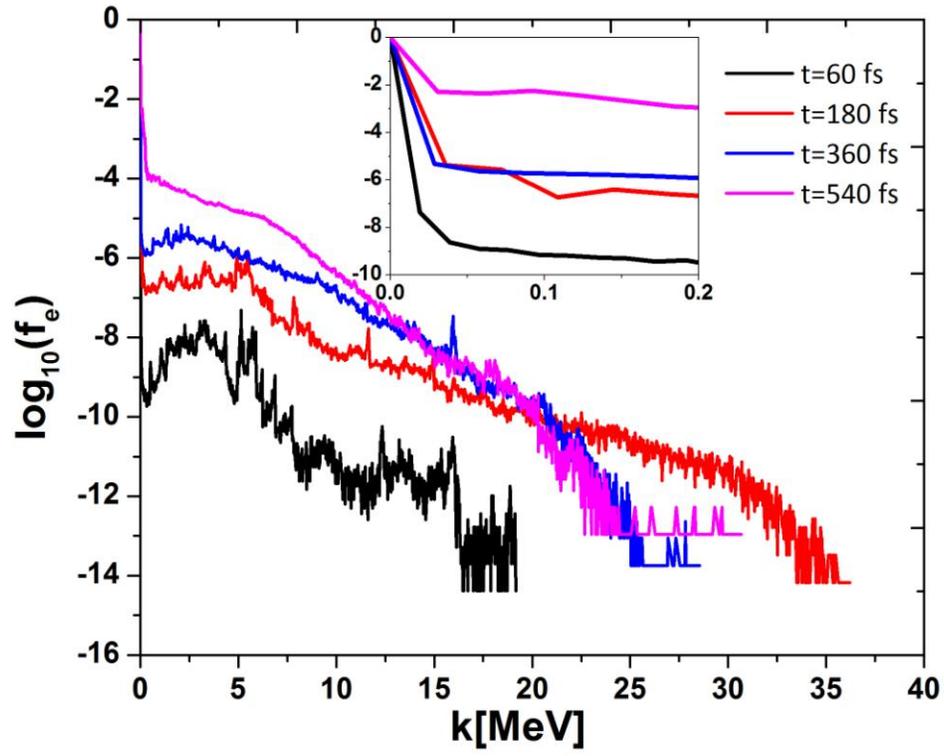

**FIG. 8. (Color online) Electron energy spectrum for parameters same as Fig. 3 at four different times. The inset is used to magnify plot in the selected range.**

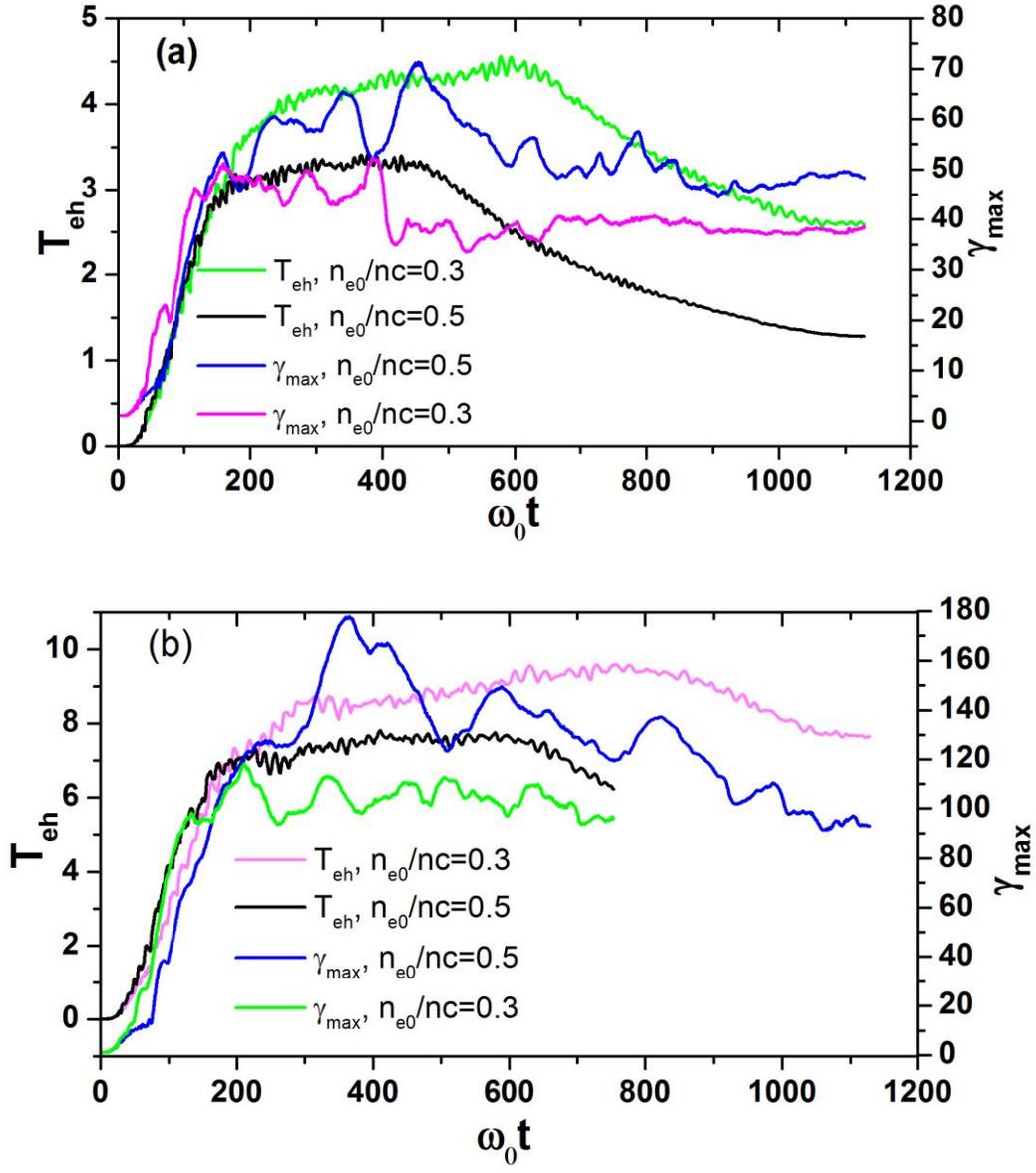

**FIG. 9. (Color online) The history of hot electron temperature and maximum gamma factor at two different densities** $n_{e0}/n_c = 0.3, 0.5$ **for** $a_0 = 6$ **(a) and** $a_0 = 10$ **(b). For all cases** $\tau_L = 100 fs$.

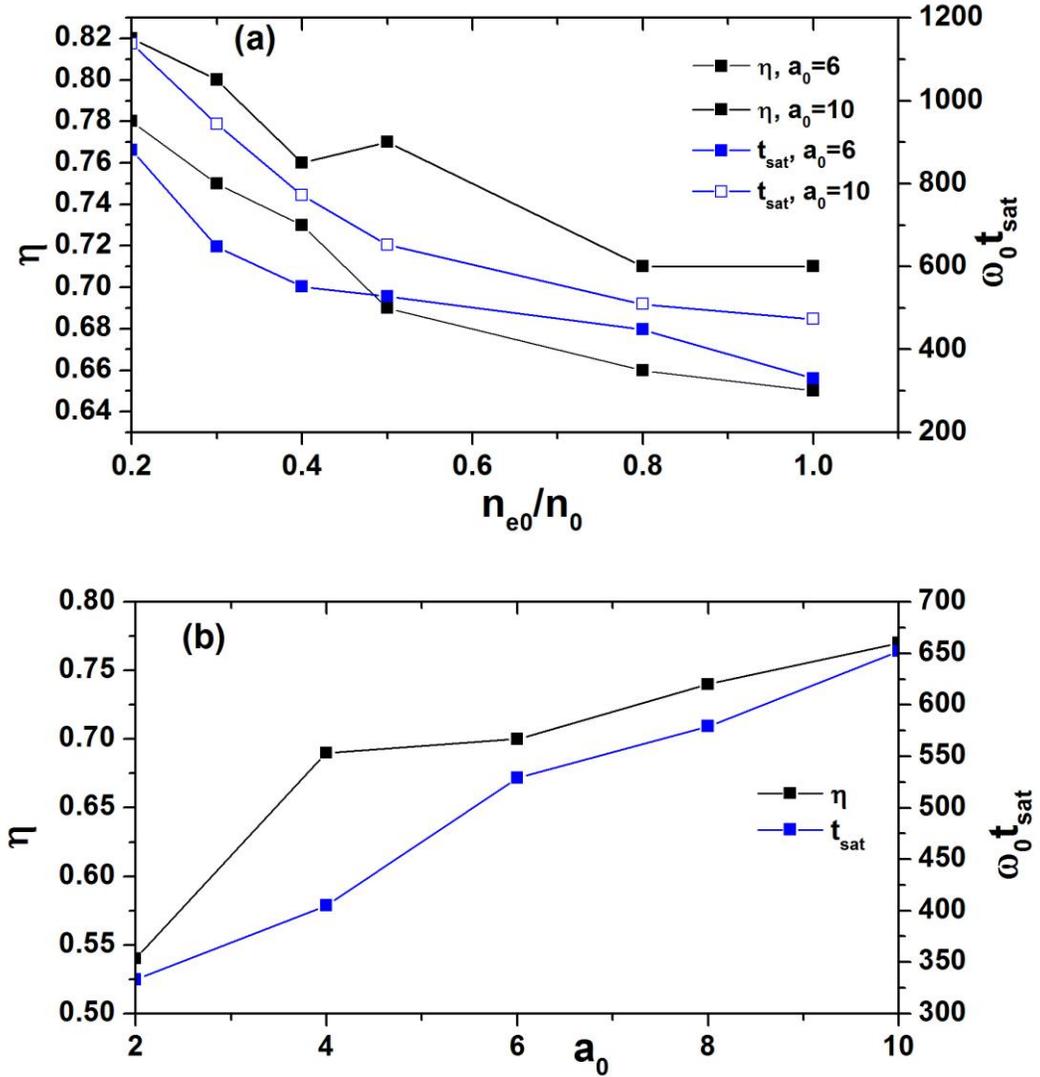

FIG. 10. (Color online) Trends of energy absorption (right) and its saturation time (left) versus plasma density for $a_0 = 6$ and $a_0 = 10$ (a), and same trends versus the laser intensity for $n_{e0}/n_c = 0.5$ (b). For all cases $\tau_L = 100\,fs$. The joint lines are for eye-guiding.

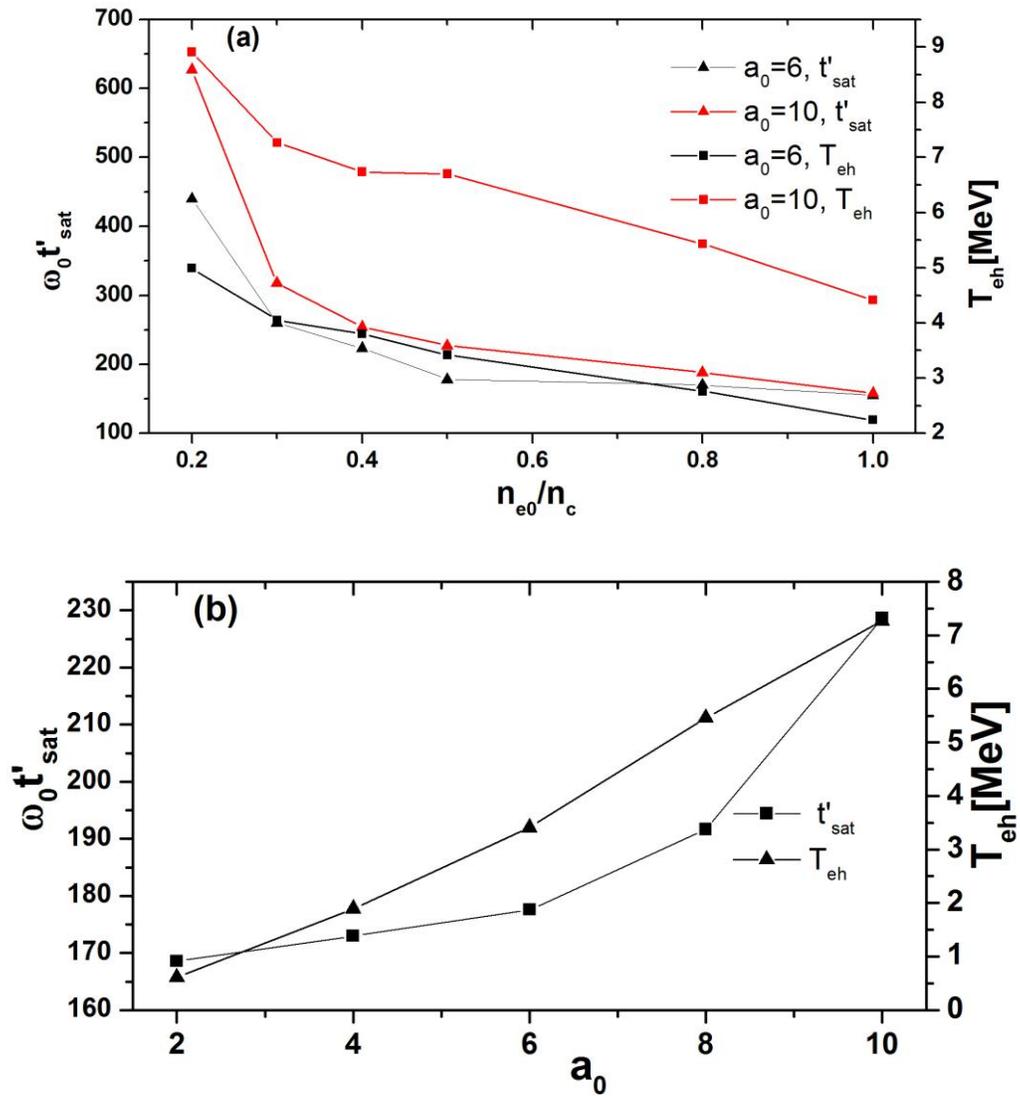

FIG. 11. (Color online) Trends of electron temperature (right) and its saturation time (left) versus plasma density (a) and laser intensity (b) at same conditions as figure 10. The joint lines are for eye-guiding.

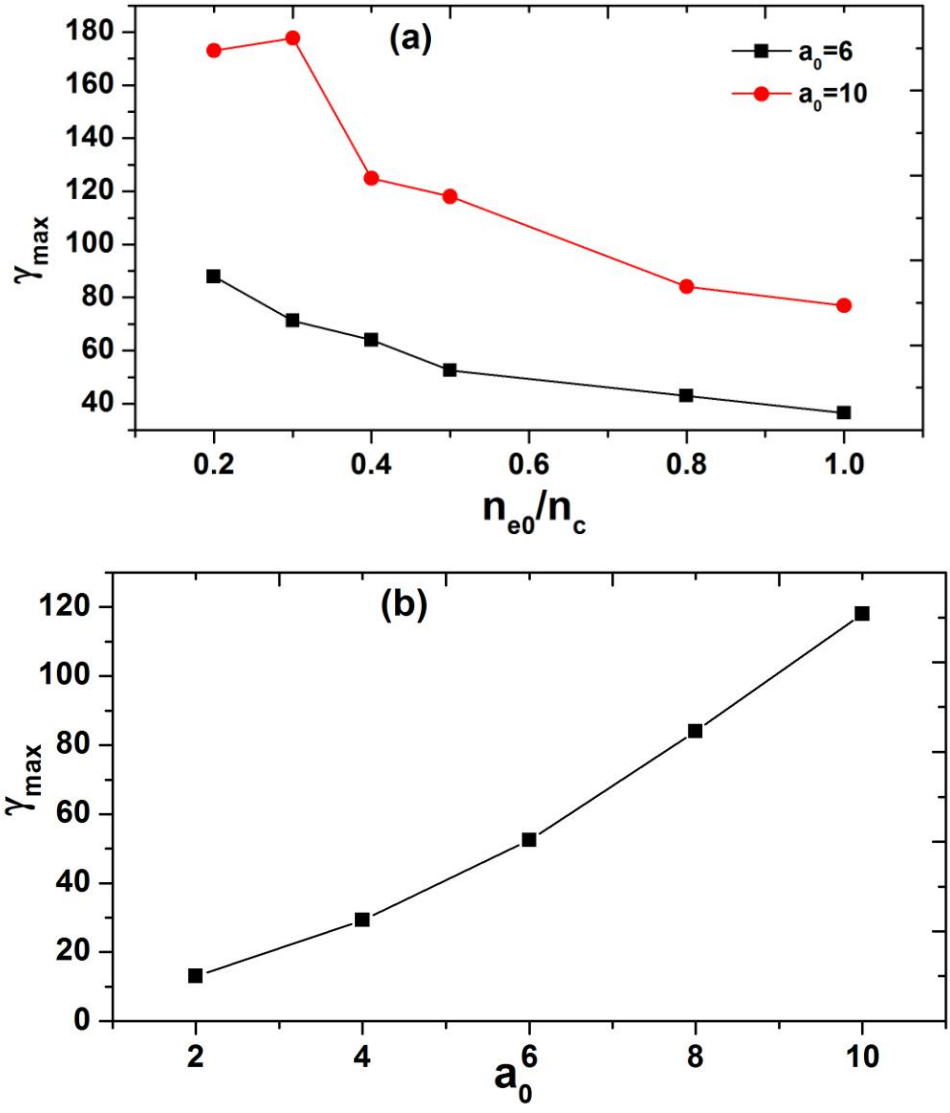

FIG. 12. (Color online) Trends of the maximum electron energy versus plasma densities (a) and laser intensities (b) at same conditions as figure 10. The joint lines are for eye-guiding.